\documentstyle[12pt,epsf]{elsart}
\newcommand{\be}{\begin{equation}}
\newcommand{\ee}{\end{equation}}
\newcommand{\bea}{\begin{eqnarray}} 
\newcommand{\eea}{\end{eqnarray}}

\newcommand{\bm}[1]{\mbox{\boldmath $#1$}}
\newcommand{\bms}[1]{\mbox{\scriptsize \boldmath $#1$}}
\newcommand{\rb}{\bm{r}}
\newcommand{\kb}{\bm{k}}
\newcommand{\qb}{\bm{q}}
\newcommand{\Rb}{\bm{R}}
\newcommand{\kd}{\kappa_{\rm D}}

\begin{document}

\begin{frontmatter}
\title{\bf The one-component plasma: a conceptual approach}
\author{M.N. Tamashiro\thanksref{tamashiro},}
\author{Yan Levin\thanksref{levin}} and  
\author{Marcia C. Barbosa}
\thanks[tamashiro]{Present address: Materials Research Laboratory, University of
California at Santa Barbara, 93106-5130 Santa Barbara (CA), USA;
e-mail: mtamash@mrl.ucsb.edu.}
\thanks[levin]{Corresponding author, e-mail: levin@if.ufrgs.br.}
\address{Instituto de F\'{\i}sica,
Universidade Federal do Rio Grande do Sul,
 Caixa Postal 15051, 91501-970 Porto Alegre (RS), Brazil}
\begin{keyword}
One-component plasma, electrolytes, structure factors, Debye-H\"uckel
theory.
\end{keyword}

\begin{abstract}
The one-component plasma (OCP) represents the simplest statistical mechanical
model of a Coulomb system. For this reason, it has been extensively 
studied over the last forty years. The advent of the integral equations has 
resulted in a dramatic improvement in our ability to carry out numerical
calculations, but came at the expense of a physical insight gained in 
a simpler analytic theory. In this paper we present an extension 
of the Debye-H\"uckel (DH) theory to the OCP. 
The theory allows for analytic 
calculations of all the thermodynamic functions, as well as the structure
factor. The theory explicitly satisfies the Stillinger-Lovett and,
for small couplings, the compressibility sum rules, 
implying its internal self consistency.
\end{abstract}
\end{frontmatter}

\section{Introduction}

The classical one-component plasma (OCP) is an idealized system of 
$N$ identical point-particles of charge $q$, in a uniform  
neutralizing background of dielectric constant $D$ \cite{Salpeter,Baus}.  
For concreteness we shall suppose that the particles are positively
 charged, while the background is negative.  Each ion, 
inside the volume $V$, is assumed to interact with the others 
exclusively through the Coulomb  potential.  The OCP has been 
extensively studied because it serves as the simplest possible 
model for a variety
of important physical systems, ranging from electrolytes and 
charge-stabilized colloids \cite{Alexander} to 
the dense stellar matter \cite{Rogers}.  
With the advent of powerful computers
 and  new developments in the liquid-state theory our ability 
to perform thermodynamic calculations on this model has seen a 
dramatic improvement characterized, in particular, by the quantitative 
agreement between the integral-equations based theories of the OCP 
and the Monte Carlo (MC) simulations \cite{Ng}.  
Unfortunately, the intrinsic complexity of 
the integral equations, instead of clarifying the underlying physics,
 tends to obscure it.  This should be contrasted with the simplicity
 and the transparency of the Debye-H\"uckel (DH) theory \cite{Debye},
 which allows for a very clear physical picture of an ionic 
solution. 

Let us recall the early history of electrolyte solutions. The first modern 
ideas about electrolytes can be traced to the pioneering work of 
the swedish physicist Svante Arrhenius \cite{Arrhenius}
at the end of the last century.  In particular, it 
was Arrhenius who realized that when salts and acids are dissolved 
in a polar solvent, the molecules become dissociated, producing cations 
and anions.  Arguing from what can now be called mean-field point of
 view, Arrhenius concluded that, since the electrolyte solution is 
charge neutral and the ions are uniformly distributed, the average 
force acting on each particle is null.  All the nontrivial 
characteristics of the electrolytes Arrhenius attributed to
 the incomplete dissociation of the molecules.  In this simple picture  
an electrolyte is treated as an ideal gas composed of three species, 
cations, anions, and  neutral molecules, whose densities are 
controlled by the law of mass action.  All the electrostatic 
interactions are neglected {\it except} in as far as treating the 
cations and the anions as distinct entities.  This simple theory has 
proven to work quite well for what are now known as weak electrolytes, 
such as Br\o{}nsted acids and bases.  In the case of strong electrolytes,
 such as NaCl or HCl, the theory proved seriously flawed.  It took
almost forty years before a satisfactory solution could be found.  It 
appeared in the form of the now famous DH theory of 
strong electrolytes \cite{Debye}.  
The great insight of Debye and H\"uckel was to 
realize that although the ions are {\it on average} uniformly distributed 
inside the solution, due to the long-range Coulomb force there exist strong 
correlations in the positions of the positively and the negatively 
charged particles;
evidently in the vicinity of a positive ion there will be an
 excess of negative particles and vice-versa.  

To make this idea concrete and to see how it applies to the OCP, let us 
fix one mobile ion at the origin
and ask what is the induced electrostatic potential in its surrounding. 
 Clearly this must satisfy the Poisson equation, 
\be
\nabla^2 \phi(\rb) =-\frac{4\pi}{D} \varrho(\rb).
\ee
To find the closure to this equation, we shall follow DH and suppose
 that the rest of the mobile ions arrange themselves in accordance 
with the Boltzmann distribution,
\be
\varrho(\rb)=q\bar\rho_+ \exp\left[-\beta q\phi(\rb)\right]-q\bar\rho_-,
\ee
where $\bar\rho_+=N/V$ is the average density of the mobile ions,
$\bar\rho_-$ is the density of the uniform neutralizing background, and 
$\beta=1/k_{\rm B}T$. The next step 
of the DH theory is to linearize the exponential factor, leading to  
\be
\varrho(\rb)=-\frac{D \kd^2}{4\pi} \phi(\rb),
\ee
where $\kd=\sqrt{4\pi\lambda_{\rm B}\bar\rho_+}$ and 
$\lambda_{\rm B}=\beta q^2/D$ are the inverse Debye screening
and the Bjerrum lengths, respectively.  Clearly, the linearization can
 only be justified in the high-temperature (weak-coupling) limit.  
The resulting Helmholtz equation, 
$\nabla^2 \phi(\rb) =\kd^2 \phi(\rb)$, can be easily integrated to 
produce a potential of the Yukawa form. The fundamental lesson of
DH is that the ions arrange themselves in such a way as to screen 
the long-range Coulomb interaction. It is this renormalization of the
interaction potential that is responsible for the existence of the 
thermodynamic limit for Coulomb systems. However, 
not everything is rosy with this simple theory.  It is sufficient to 
look at the charge-density distribution,
\be
\varrho(\rb)=-\frac{q\kd^2}{4\pi r}\,{\rm e}^{-\kd r},
\ee
to notice that something 
went seriously wrong.

Clearly, the physical restriction that $\varrho(\rb)\ge -q\bar\rho_-$ is
strongly violated in the region near the fixed ion. 
The problem can be traced 
back to the linearization of the Boltzmann factor, which is
unjustified at short distances, since there the electrostatic potential is 
not small even for high temperatures. Fortunately, not all is lost. 
A simple solution to overcome this difficulty was suggested by 
Nordholm \cite{Nordholm}, who proposed an augmentation of the DH theory 
to include an effective spherical cavity around the fixed 
ion, inside which no other ions can penetrate.  The presence of such 
a cavity is quite reasonable, since the electrostatic repulsion between 
the like-charged ions should prevent them from coming into close contact.  
Furthermore, we can estimate the size of the hole, $h$, by comparing 
the repulsive Coulomb energy with the kinetic thermal energy, $q^2/Dh 
\sim k_BT$, or $h \sim q^2/Dk_BT$.  Evidently, the higher the temperature,
 the smaller the size of the exclusion region.  This, of course, is
intuitive, since at higher temperatures the ions will have 
more kinetic energy to overcome the mutual repulsion.  
A more consistent way of 
defining the hole size $h$ is to require an overall charge neutrality \cite{Nordholm}, 
or, equivalently,
 the continuity of the potential and of the electric field across
 the hole boundary.  Performing a simple calculation, we find
\bea
h&=&d\left[\omega(\Gamma)-1\right]/\sqrt{3\Gamma}, 
\label{eqn:ocp_hole} \\
\omega(\Gamma)&=&\left[1+\left(3\Gamma \right)^{3/2}\,\right]^{1/3},
\eea
where we have defined the usual coupling constant for the OCP,
$\Gamma=\beta q^2/Dd$, and $d$ is the characteristic
length scale, $d =\left(4 \pi \bar\rho_+^3/3\right)^{-1}$.  
In the low-coupling limit this reduces to the 
energetically determined expression for the size of the cavity,
$h\approx\lambda_{\rm B}$. Nordholm 
was able to demonstrate that this Debye-H\"uckel plus hole (DHH) theory 
produces an equation of state for the OCP which is in good agreement 
with the MC simulations.  The question, however, still remains 
to what extent the hole is a physical object or just a convenient 
mathematical trick to correct for the linearization of the
 Poisson-Boltzmann (PB) equation.  Clearly, if the cavity postulated  
by the DHH theory is real, the best way to study it is by considering 
the structure factor.  In particular, if everything is all right with the 
DHH theory, the structure factor obtained on its basis should be in good 
agreement with the MC simulations.  Unfortunately, it is well 
known that the traditional ways of obtaining the correlation functions 
out of the DH theory lead to expressions which are seriously 
flawed \cite{Lee}. In the case of the restricted primitive model 
(RPM), equisized spheres carrying charges $\pm q$, the correlation 
functions violate the well-known Stillinger-Lovett sum 
rules \cite{Stillinger} and do not reproduce the charge-density
oscillations known to exist at high densities \cite{Kirkwood}.  

Recently Lee and Fisher \cite{Lee}  have proposed an extension of
 the DH theory of strong electrolytes 
to nonuniform densities.  The generalized DH theory 
(GDH) allows the calculation of the density-density and of the charge-charge 
correlation functions in a most natural way, through a functional  
differentiation of the free-energy functional.  Furthermore, since the 
theory is constructed at the level of free energy, it is internally 
self consistent, as can be judged by the various sum rules that it
 satisfies.  This, of course, is a great advantage over the traditional
 integral-equations based theories, which are constructed at the level of 
the correlation functions and depend on the route taken to 
the thermodynamics \cite{integral_equations}, 
i.e. virial, compressibility, etc.
 The comparison of the GDH theory
 with experiments or simulations, however, is made difficult  by the 
same flaw (or virtue, depending on how one looks at it) 
that the original DH theory suffers, its linearity.  The 
linearity of the DH theory for electrolytes is both a blessing 
and a curse.  It ensures the internal self consistency of the theory, but
 is also responsible for undercounting the configurations in which the 
oppositely charged ions come into a close contact, forming dipolar pairs. 
 It was shown recently how this difficulty can be overcome in the context
 of the Debye-H\"uckel-Bjerrum (DHBj) theory \cite{Fisher}, by allowing
for the existence of a chemical 
equilibrium between the monopoles and dipoles. 
Unfortunately this stratagem is difficult to implement in the case of 
the GDH theory. The goal of this paper, then, is two-fold: 
test the physical nature of the cavity in the DHH theory and,
 by apply the GDH theory to the OCP --- 
which is free from the cluster formations
 that plague RPM --- test the extent of its validity.

\section{The generalized Debye-H\"uckel theory for the 
one-component plasma}

The DHH theory for the OCP is extended to allow for a 
nonuniform, slowly varying ionic density, 
\be
\rho_{+}(\rb) = \bar\rho_+\left(1+\Delta\cos \kb\cdot\rb\right).
\label{eqn:rho_variation}
\ee
The negative background, as in the original OCP theory,
is maintained uniform,
\be
\rho_{-}(\rb) = \bar\rho_-,\quad\forall \rb.
\ee
To preserve the electroneutrality on long-length scales, the 
overall equilibrium densities must be equal, $\bar\rho_+=\bar\rho_-$.

The Helmholtz free-energy, $F$, is a functional of the ionic density  
$\rho_{+}(\rb)$.  The direct correlation function, 
$C_{++}(\rb_1-\rb_2)$, is defined in terms of the second derivative of the 
excess free energy,
\bea
C_{++}(\rb_1-\rb_2) &\equiv& -\beta\left.\frac{\delta^2 \left\{
F\left[\rho_{+}(\rb)\right] - F_{\rm ideal}\left[\rho_{+}(\rb)\right]
\right\}}
{\delta\rho_{+}(\rb_1)\,\delta\rho_{+}(\rb_2) }
\right|_{\rho_{+}(\bms{r})=\bar\rho_{+}} \nonumber \\
&=& \frac{\delta(\rb_1-\rb_2)}{\bar\rho_{+}}
-\beta \left.\frac{\delta^2 F\left[\rho_{+}(\rb)\right] }
{\delta\rho_{+}(\rb_1)\,\delta\rho_{+}(\rb_2) }
\right|_{\rho_{+}(\bms{r})=\bar\rho_{+}}. 
\label{eqn:c_2}
\eea
Here, $F_{\rm ideal}\left[\rho_{+}(\rb)\right]$ 
is the usual ideal-gas Helmholtz 
free-energy functional,
\be
\beta F_{\rm ideal}\left[\rho_{+}(\rb)\right]=
\int {\rm d}^3\rb'\, \rho_{+}(\rb')
\left\{ \ln\left[ \rho_{+}(\rb') \Lambda^3\right] -1 \right\},
\label{eqn:ideal}
\ee
where $\Lambda$ is the thermal de Broglie wavelength.

The direct correlation function is connected with the total
correlation function, $H(\rb)$, through the Ornstein-Zernike relation,
\be
 H(\rb) = C_{++}(\rb) + \bar\rho_+ 
\int {\rm d}^3\rb'\; C_{++}(\rb-\rb')\,H(\rb'),
\ee
which in the Fourier space can be written as  
\be
\hat{H}(\kb) = \frac{\hat{C}_{++}(\kb)}
{1-\bar\rho_+\hat{C}_{++}(\kb)} ,
\ee
where $\hat{C}_{++}(\kb)$ and $\hat{H}(\kb)$ are 
the Fourier transforms of the direct and the total correlation functions,
respectively,   
\bea
\hat{C}_{++}(\kb) &=& 
\int {\rm d}^3\rb\; C_{++}(\rb) \exp \left({\rm i} \kb\cdot\rb\right), \\
\hat{H}(\kb) &=& 
\int {\rm d}^3\rb\; H(\rb) \exp \left({\rm i} \kb\cdot\rb\right).
\eea
The structure factor is defined as 
\be
S(\kb) \equiv 1 + \bar\rho_+\hat{H}(\kb)=
\frac{1}{1-\bar\rho_+\hat{C}_{++}(\kb)}. \label{eqn:sk}
\ee
Evidently, the knowledge of the direct correlation function
$C_{++}(\rb)$ is equivalent
to the knowledge of the structure factor $S(\kb)$.
 
To proceed, we impose an infinitesimal variation on the mobile-ion 
density, Eq.~(\ref{eqn:rho_variation}),  
and expand the reduced Helmholtz 
free-energy functional density,
$f \equiv \beta F/V$, in powers of $\Delta$. 
To second order, the variation $\delta f$ 
can be written as (see details in appendix B)
\be
\delta{f}\left[\rho_{+}(\rb)\right]
\equiv  f\left[ \rho_{+}(\rb)\right] - 
{f} \left[ \bar\rho_{+} \right]  =
\beta\bar\mu \bar\rho_+ \Delta \delta_{{\bms{k}}{\bf 0}} + 
\frac14 {S}^{-1}(\kb) \bar\rho_+ \Delta^2 
\left(1+\delta_{{\bms{k}}{\bf 0}}\right) , 
\label{eqn:f_variation}
\ee
where $\beta\bar\mu=\partial f/\partial\bar\rho_+$ 
is the equilibrium chemical potential,  
$\delta_{{\bms{k}}{\bf 0}} = 
{\frac{1}{V}}(2\pi)^3 \delta^3(\kb)$
is the Kronecker delta and $\delta^3(\kb)$ is the
three-dimensional Dirac delta function.
The free-energy density of the homogeneous reference system, 
$f\left[ \bar\rho_{+} \right]$, is obtained by setting $\Delta=0$ in
the expression for $f\left[ \rho_{+}(\rb) \right]$. Clearly, if we 
are able to construct the free-energy functional for a nonuniform 
system, the structure factor, $S(\kb)$,  
can be read directly from the second-order term. 

We proceed in a way exactly analogous to the usual DH theory. Let us fix
one positive 
ion at $\rb'$ and ask what is the electrostatic potential,
$\phi(\rb,\rb')$, at a point $\rb$ in its surrounding. We shall assume that, 
just like in the uniform case, this potential satisfies the PB equation, 
\bea
\nabla_r^2 \phi(\rb,\rb') &=& 
-\frac{4\pi q}{D} \left\{\delta^3 (\rb-\rb') + 
\rho_{+}(\rb) \exp\left[-\beta q \tilde\phi(\rb,\rb') \right] 
-\rho_{-}(\rb)  \right\} . \label{eqn:pb_equation}
\eea
A crucial point to remark \cite{Lee} is that the potential 
which appears in the Boltzmann factor of~(\ref{eqn:pb_equation}), 
$\tilde\phi(\rb,\rb')$, represents a 
\textit{local-induced} potential,
\be
\tilde\phi(\rb,\rb') \equiv \phi(\rb,\rb') - \Phi(\rb),
\ee
obtained by extracting from the total potential $\phi(\rb,\rb')$
an \textit{imposed} electrostatic
potential, $\Phi(\rb)$, produced by the neutralizing background and 
the imposed charge-density variation~(\ref{eqn:rho_variation}),
\be
\nabla^2 \Phi(\rb)
= -\frac{4\pi q}{D} \left[\rho_{+}(\rb)-\bar\rho_{-}\right]
= -\frac{4\pi q\bar\rho_{+}}{D} \Delta\cos \kb\cdot\rb, 
\qquad\forall\rb. \label{eqn:poisson}
\ee

With the separation of the total potential $\phi(\rb,\rb')$ 
into two parts, the Helmholtz free-energy functional can be 
written as 
\be
F\left[\rho_{+}(\rb) \right] =
F_{\rm ideal}\left[\rho_{+}(\rb) \right] +
F_{\rm imposed}\left[\rho_{+}(\rb) \right] +
F_{\rm induced}\left[\rho_{+}(\rb) \right] ,
\ee
where the excess free energies are obtained through the Debye 
charging process \cite{Debye,Lee},
\bea
F_{\rm imposed}\left[\rho_{+}(\rb) \right] &=&
q \int {\rm d}^3\rb'\,\left[\rho_+(\rb')-\bar\rho_-\right]
\int\limits_{0}^{1} {\rm d}\lambda \,\Phi(\rb',\lambda q) , \\
F_{\rm induced}\left[\rho_{+}(\rb) \right] &=&
q \int {\rm d}^3\rb'\,\rho_+(\rb') 
\int\limits_{0}^{1} {\rm d}\lambda \,\psi(\rb',\lambda q).
\label{eqn:f_induced} 
\eea
In Eq.~(\ref{eqn:f_induced}),
$\psi(\rb')$ is the \textit{mean induced} electrostatic 
potential felt by the positive ion fixed at $\rb'$,
\be
\psi(\rb') \equiv \lim_{\bms{r}\to\bms{r}'} 
\left[\tilde\phi(\rb,\rb')-\frac{q}{D|\rb-\rb'|} \right]. 
\ee
 
Linearization of the Boltzmann factor of~(\ref{eqn:pb_equation}) 
results in the 
GDH equation for the induced 
potential,
\bea
\nabla_r^2 \tilde\phi(\rb,\rb') &=& -\frac{4\pi q}{D}
\left[\delta^3 (\rb-\rb') - \rho_{+}(\rb)\right] \nonumber\\
&=& -\frac{4\pi q}{D}
\left[\delta^3 (\rb-\rb') - \bar\rho_{+}(1+\Delta\cos\kb\cdot\rb)\right],
\quad\mbox{ for } |\rb-\rb'| \leq h, 
\label{eqn:induced_inside} \\
\nabla_r^2 \tilde\phi(\rb,\rb') &=& 
\frac{4\pi\beta q^2\rho_{+}(\rb)}{D}\, \tilde\phi(\rb,\rb') \nonumber\\
&=& \kd^2 \left(1+\Delta\cos\kb\cdot\rb\right) \tilde\phi(\rb,\rb'),
\quad\mbox{ for } |\rb-\rb'| \geq h. \label{eqn:induced_outside}
\eea
As discussed in the introduction, to prevent the unphysical artifacts 
of the linearization of the PB equation, we have explicitly introduced a 
cavity $h$ around the fixed ion at $\rb'$, given 
by~Eq.~(\ref{eqn:ocp_hole}), into which no other mobile ions can penetrate. 

In the following subsections we shall obtain the 
contributions to the variation of the free-energy density. 

\subsection{The ideal-gas contribution} 

The ideal-gas contribution is given by~(\ref{eqn:ideal})
with the imposed mobile-ion charge
distribution~(\ref{eqn:rho_variation}).
Expanding~(\ref{eqn:ideal}) up to order $\Delta^2$ 
and using the integrals~(\ref{eqn:coskr}) and~(\ref{eqn:coskr2}), 
we obtain the ideal-gas contribution to the variation of the
reduced free-energy density,
\be
\delta{f}_{\rm ideal}\left[\rho_{+}(\rb)\right]=
\ln\left(\bar\rho_+ \Lambda^3\right) \! \bar\rho_+ \Delta 
\delta_{{\bms{k}}{\bf 0}}
+ \frac14\bar\rho_+ \Delta^2 \left(1+\delta_{{\bms{k}}{\bf 0}} \right).
\ee
 
\subsection{The imposed electrostatic potential contribution}

The imposed electrostatic potential satisfies the Poisson 
equation~(\ref{eqn:poisson}), 
whose formal solution can be written as 
\be
\Phi(\rb) = \frac{q\bar\rho_+}{D}\Delta
\int{\rm d}^3\rb'\; \frac{\cos \kb\cdot\rb'}{|\rb-\rb'|}.
\ee
The contribution to the Helmholtz free-energy functional 
is obtained through the Debye charging process \cite{Debye},
\bea
{F_{\rm imposed}}\left[ \rho_+(\rb) \right] &=& 
q \int {\rm d}^3\rb'\,\left[\rho_+(\rb')-\bar\rho_-\right]
\int\limits_{0}^{1} {\rm d}\lambda \,\Phi(\rb',\lambda q)\nonumber\\
&=& 
\frac{q^2 \bar\rho_+^2}{D}\Delta^2  
\int {\rm d}^3\rb\, {\rm d}^3\rb' 
\, \frac{\cos \kb\cdot\rb \: 
\cos \kb\cdot\rb' }{|\rb-\rb'|}
\int\limits_{0}^{1} {\rm d}\lambda \,\lambda .
\eea
In this case the charging merely produces a trivial factor of 
$1/2$ and using the 
integral~(\ref{eqn:coskr_overr2}),
we obtain the contribution of the imposed 
electrostatic potential to the variation of the 
reduced free-energy density,
\be
\delta f_{\rm imposed}\left[ \rho_+(\rb) \right]=
\frac14 \left(\frac{\kd}{k}\right)^{\!2}\! 
\bar\rho_+\Delta^2\!
 \left( 1+\delta_{{\bms{k}}{\bf 0}} \right) .
\ee

\subsection{The induced electrostatic potential contribution}
 
The induced electrostatic potential satisfies the 
GDH equation, given
by~(\ref{eqn:induced_inside})~and~(\ref{eqn:induced_outside}).
It is convenient to rewrite them in a spherical coordinate 
system centered on the
positive ion fixed at $\rb'$. Introducing the difference vector
\be
\Rb = \rb - \rb',
\ee
the GDH equation for the induced potential reads
\be
\nabla_R^2 \tilde\phi(\Rb+\rb',\rb')=
\left\{
\begin{array}{cl}
-\frac{\displaystyle 4\pi q}{\displaystyle D}
\left\{\delta^3 (\Rb) - \bar\rho_{+}\left[1+\Delta\cos\kb\cdot
\left(\Rb+\rb'\right)\right]\right\},
 & \qquad \mbox{ for } |\Rb| \leq h, \\
\kd^2 \left[1+\Delta\cos\kb\cdot\left(\Rb+\rb'\right)\right]
\tilde\phi(\Rb+\rb',\rb'),
& \qquad \mbox{ for } |\Rb| \geq h.
\label{eqn:induced_potential1}
\end{array} 
\right.
\ee
Using the Green's function $G\left(\Rb,\Rb'\right)$ associated 
with~(\ref{eqn:induced_potential1}) derived in 
appendix C, it can be transformed 
into the integral equation
\be
\tilde\phi(\rb,\rb')=
\tilde\phi(\Rb+\rb',\rb')=
\frac{1}{D}\int {\rm d}^3 \Rb'\, \varrho\left(\Rb'\right)
G\left(\Rb,\Rb'\right),
\ee
where the effective charge density,
 $\varrho\left(\Rb\right)$, is given by
\bea
\varrho\left(\Rb\right)&=&
\left\{
\begin{array}{cl}
q\delta^3(\Rb) - q\bar\rho_{+}\left[1+\Delta\cos\kb\cdot(\Rb+\rb')\right],
 & \qquad \mbox{ for } R \leq h, \\
-{\displaystyle\frac{D}{4\pi}}\kd^2\tilde\phi\left(\Rb+\rb',\rb'\right)
 \Delta\cos\kb\cdot(\Rb+\rb'),
& \qquad \mbox{ for } R \geq h.
\end{array} 
\right.
\label{eqn:nonhomogeneity}
\eea
This equation can be solved perturbatively in powers 
of $\Delta$ (see appendix D).

The mean induced electrostatic potential felt by the
positive ion fixed at $\rb'$,
\be
\psi(\rb')=
 \lim_{{\bms{R}}\to{\bf 0}}\left[\tilde\phi(\Rb+\rb',\rb')
-\frac{q}{DR}\right],
\ee
can be written, to order $\Delta^2$, as (see derivation in appendix D)
\bea
\beta q\psi(\rb')&=& -\frac12 x (x+2) 
-\Delta\cos\kb\cdot\rb' \left[\frac{1}{\alpha^2} 
-\frac{\sin \alpha x}{\alpha^3(1+x)}-
\frac{\cos \alpha x}{\alpha^2(1+x)} + \frac{x}{1+x}\,
{\cal I}_0^+(x,\alpha) \right]\nonumber\\
&&\mbox{}\!+ \frac{1}{\alpha(1+x)}
\Delta^2\sum_{\ell=0}^{\infty} \left(2\ell+1\right)
\cos^2\left(\kb\cdot\rb'+\ell\frac{\pi}{2}\right) 
\frac{x^{\ell+2} j_{\ell+1}(\alpha x)}{g_{\ell+1}(x)}\,
{\cal I}_\ell^+(x,\alpha)\nonumber\\
&&\mbox{}\!+  \frac{x}{1+x}
\Delta^2\sum_{\ell=0}^{\infty} \left(-1\right)^{\ell+1} \left(2\ell+1\right)
\cos^2\left(\kb\cdot\rb'+\ell\frac{\pi}{2}\right)\nonumber\\
&&\mbox{}\!\times
\left\{\frac12 \frac{g_{\ell+1}(-x)}{g_{\ell+1}(x)}
\left[{\cal I}_\ell^+(x,\alpha)\right]^2+
{\cal I}_\ell^+(x,\alpha){\cal I}_\ell^-(x,\alpha)
-{\cal I}_\ell^0(x,\alpha)\right\}.
\eea
where $x=\kd h$, $\alpha=k/\kd$, $j_\ell(\xi)$ is the spherical Bessel function
of the first kind, 
\be
j_\ell(\xi) = \sqrt{\frac{\pi}{2 \xi}}\, J_{\ell+1/2} (\xi),
\ee
$g_\ell(\xi)$ is the $\ell$-th grade polynomial
associated with the modified spherical
Bessel function of the third kind, $k_\ell(\xi)$,
\be
g_\ell(\xi)\equiv {\rm e}^{\xi}{\xi^{\ell+1}}k_\ell(\xi)
= \sum_{m=0}^\ell\frac{\Gamma(\ell+m+1)}{ 2^m m!\,
\Gamma(\ell-m+1)}\,\xi^{\ell-m} = 
\sum_{m=0}^\ell
\frac{(2m)!}{2^m m!}\, {{\ell+m}\choose{2m}}\,\xi^{\ell-m} ,
\ee
where $\Gamma(m)=(m-1)!$ is the Euler gamma function,  
and $\{{\cal I}_\ell^\nu\}, \nu=\pm,0,$ are the 
one-dimensional quadratures
\bea
{\cal I}_\ell^-(s,\alpha) &=& 
\int\limits_0^{s} {\rm d}\xi\, \xi^{-\ell} g_\ell(-\xi)
j_\ell(\alpha \xi), \\
{\cal I}_\ell^0(x,\alpha) &=& 
\int\limits_x^\infty {\rm d}s\, s^{-\ell} g_\ell(s)
j_\ell(\alpha s){\cal I}_\ell^-(s,\alpha)\exp\left[2(x-s)\right], \\
{\cal I}_\ell^+(x,\alpha) &=& 
\int\limits_x^\infty {\rm d}s\, s^{-\ell} g_\ell(s)
j_\ell(\alpha s)\exp\left[2(x-s)\right].
\eea

The contribution to the Helmholtz free-energy functional 
is obtained through the Debye charging process \cite{Debye,Lee},
\bea
{F_{\rm induced}}\left[ \rho_+(\rb) \right]&=&
q \int {\rm d}^3\rb'\, \rho_+(\rb')
\int\limits_{0}^{1} {\rm d}\lambda 
\,\psi(\rb',\lambda q)  \nonumber\\
&=& q \bar\rho_+\int {\rm d}^3\rb'\,
\left(1+\Delta\cos\kb\cdot\rb'\right)
\int\limits_{0}^{1} {\rm d}\lambda 
\,\psi(\rb',\lambda q),
\eea
which yields the reduced free-energy density,
\be
f_{\rm induced}\left[\rho_+(\rb)\right] = \bar\rho_+f_0 +
\left[f_0+\frac12 f_1(\alpha)\right]\bar\rho_+\Delta \delta_{{\bms{k}}{\bf 0}}
+ \frac14\left[f_1(\alpha)+f_2(\alpha)\right]\bar\rho_+\Delta^2 
\left( 1+\delta_{{\bms{k}}{\bf 0}} \right),
\ee
from which the variation follows,
\bea
\delta f_{\rm induced}\left[\rho_+(\rb)\right]&=&
f_{\rm induced}\left[\rho_+(\rb)\right] 
-f_{\rm induced}\left[\bar\rho_+\right] \nonumber\\
&=&
\left[f_0+\frac12 f_1(\alpha)\right]\bar\rho_+\Delta\delta_{{\bms{k}}{\bf 0}}
+ \frac14\left[f_1(\alpha)+f_2(\alpha)\right]\bar\rho_+\Delta^2 
\left( 1+\delta_{{\bms{k}}{\bf 0}} \right),
\eea
where 
\bea
f_0&=& -\frac12 
\int\limits_0^1 \frac{{\rm d}\lambda}{\lambda}
\, x_\lambda(x_\lambda+2) = 
 -\frac12\int\limits_1^\omega {\rm d}\omega_\lambda\,
\frac{\omega_\lambda^2\left(\omega_\lambda+1\right)}
{\omega_\lambda^2+\omega_\lambda+1} \nonumber\\
&=& \frac14 \left[ 1-\omega^2+\frac{2\pi}{3\sqrt{3}}
+ \ln\left(\frac{\omega^2+\omega+1}{3}\right)
-\frac{2}{\sqrt{3}}\tan^{-1}\left(\frac{2\omega+1}{\sqrt{3}}\right)
\right], \label{eqn:f0}\\ 
f_1(\alpha)&=&-\frac1{\alpha^2}
+ \frac{2}{\alpha^3}\int\limits_0^1
{\rm d}\lambda\,\frac{\lambda^2}{\omega_\lambda}
\sin(\alpha x_\lambda/\lambda)
+ \frac{2}{\alpha^2}\int\limits_0^1
{\rm d}\lambda\,\frac{\lambda}{\omega_\lambda}
\cos(\alpha x_\lambda/\lambda)\nonumber\\
&&\mbox{}\! -2\int\limits_0^1 \frac{{\rm d}\lambda}{\lambda}
\,\frac{\omega_\lambda-1}{\omega_\lambda}\,
{\cal I}_0^+(x_\lambda,\alpha/\lambda), \\
f_2(\alpha)&=&
\sum_{\ell=0}^{\infty} \left(2\ell+1\right)
\left[\frac{1+(-1)^\ell\delta_{{\bms{k}}{\bf 0}} }
{1+\delta_{{\bms{k}}{\bf 0}} }\right]
\left\{ \frac{2}{\alpha}\int\limits_0^1 {\rm d}\lambda\,
\frac{x_\lambda^{\ell+2} j_{\ell+1}(\alpha x_\lambda/\lambda)}
{\omega_\lambda\, g_{\ell+1}(x_\lambda)}\,
{\cal I}_\ell^+(x_\lambda,\alpha/\lambda)\right.\nonumber\\
&&\mbox{}\!+\left(-1\right)^{\ell+1} 
\int\limits_0^1 \frac{{\rm d}\lambda}{\lambda}\,
\frac{\omega_\lambda-1}{\omega_\lambda}\, 
\left[ \frac{g_{\ell+1}(-x_\lambda)}{g_{\ell+1}(x_\lambda)}
\left[{\cal I}_\ell^+(x_\lambda,\alpha/\lambda)\right]^2\right.
\nonumber\\
&& \left.\vphantom{\int\limits_0^1}\!
\left.\vphantom{\frac{g_\ell}{g_\ell}}\!+
2 {\cal I}_\ell^+(x_\lambda,\alpha/\lambda)
{\cal I}_\ell^-(x_\lambda,\alpha/\lambda)
-2{\cal I}_\ell^0(x_\lambda,\alpha/\lambda)\right] \right\} , \\
x_\lambda&=& \omega_\lambda-1, \\
\omega_\lambda&=& \left[1+\lambda^3\left(3\Gamma
\right)^{3/2}\,\right]^{1/3},\\
\omega&=& \omega_{\lambda=1}=
 \left[1+\left(3\Gamma \right)^{3/2}\,\right]^{1/3}.
\eea
We note that the reduced induced free-energy density for 
the reference system, 
\bea
f_{\rm induced}\left[\bar\rho_+\right] &=&  \bar\rho_+ f_0 \nonumber\\
&=& \frac14 \bar\rho_+\left[ 1-\omega^2+\frac{2\pi}{3\sqrt{3}}
+ \ln\left(\frac{\omega^2+\omega+1}{3}\right)
-\frac{2}{\sqrt{3}}\tan^{-1}\left(\frac{2\omega+1}{\sqrt{3}}\right)
\right],
\eea
is the same as the one previously obtained by Nordholm \cite{Nordholm}.

\section{Analytical results and the sum rules}

Collecting all the contributions to the variation of the 
reduced free-energy density,
\be
\delta f\left[\rho_+(\rb)\right] = 
\delta f_{\rm ideal}\left[\rho_+(\rb)\right]+
\delta f_{\rm imposed}\left[\rho_+(\rb)\right]+
\delta f_{\rm induced}\left[\rho_+(\rb)\right],
\ee
and comparing with the expansion given by~(\ref{eqn:f_variation}), 
we obtain the equilibrium chemical potential,
\be
\beta\bar\mu=
\ln\left(\bar\rho_+ \Lambda^3\right)
+f_0 + \frac12\lim_{\alpha\to 0} f_1(\alpha)
= \ln\left(\bar\rho_+ \Lambda^3\right)
+f_0 + \frac1{12}\left(1-\omega^2\right), 
\ee
which corresponds to the usual OCP chemical potential \cite{Nordholm},
and the structure factor,
\bea
S^{-1}(\kb)&=&
1+ \frac{2}{\alpha^3}\int\limits_0^1
{\rm d}\lambda\,\frac{\lambda^2}{\omega_\lambda}
\sin(\alpha x_\lambda/\lambda)
+ \frac{2}{\alpha^2}\int\limits_0^1
{\rm d}\lambda\,\frac{\lambda}{\omega_\lambda}
\cos(\alpha x_\lambda/\lambda)\nonumber\\ 
&&\mbox{}\!-2\int\limits_0^1 \frac{{\rm d}\lambda}{\lambda}
\,\frac{\omega_\lambda-1}{\omega_\lambda}\,
{\cal I}_0^+(x_\lambda,\alpha/\lambda)\nonumber\\
&&\mbox{}\!+ \sum_{\ell=0}^{\infty} \left(2\ell+1\right)
\left\{ \frac{2}{\alpha}\int\limits_0^1 {\rm d}\lambda\,
\frac{x_\lambda^{\ell+2} j_{\ell+1}(\alpha x_\lambda/\lambda)}
{\omega_\lambda\, g_{\ell+1}(x_\lambda)}\,
{\cal I}_\ell^+(x_\lambda,\alpha/\lambda)\right.\nonumber\\
&&\mbox{}\!+\left(-1\right)^{\ell+1} 
\int\limits_0^1 \frac{{\rm d}\lambda}{\lambda}\,
\frac{\omega_\lambda-1}{\omega_\lambda}\, 
\left[ \frac{g_{\ell+1}(-x_\lambda)}{g_{\ell+1}(x_\lambda)}
\left[{\cal I}_\ell^+(x_\lambda,\alpha/\lambda)\right]^2\right.
\nonumber\\
&& \left.\vphantom{\int\limits_0^1}\!
\left.\vphantom{\frac{g_\ell}{g_\ell}}\!+
2 {\cal I}_\ell^+(x_\lambda,\alpha/\lambda)
{\cal I}_\ell^-(x_\lambda,\alpha/\lambda)
-2{\cal I}_\ell^0(x_\lambda,\alpha/\lambda)\right] \right\} .
\label{eqn:skGDH}
\eea
This is the central result of this paper, the explicit
expression for the structure factor of the OCP, given in terms of
an infinite series. To check the internal
consistency of the theory, we explore how well it satisfies various
known sum rules. All these can be conveniently summarized in the
exact, small $\kb$ expansion of the structure factor \cite{Baus}, 
\be
S^{-1}_{\rm exact}(\kb) =
\left(\frac{\kd}{k}\right)^2 + 
\frac{\beta}{\bar\rho_+\chi}+{\cal O}(k^2), \label{eqn:sk_exact}
\ee
where $\chi$ is the compressibility of the OCP.
The first term is the result of the charge neutrality and
of the Stillinger-Lovett second-moment condition  \cite{Stillinger}, 
while the second term corresponds to the fourth-moment or the 
``compressibility'' sum rule \cite{Baus}.
The inverse compressibility
is defined thermodynamically in terms of the variation of the pressure,
\be
 P=\frac{\bar\rho_+}{\beta}
\left(1+ \bar\rho_+
\frac{\partial f_0}{\partial \bar\rho_+} \right) , 
\ee
with respect to the density, 
\be
\frac{\beta}{\bar\rho_+\chi_P} \equiv 
\beta \frac{\partial P}{\partial \bar\rho_+} =
1+ 2\bar\rho_+ \frac{\partial f_0}{\partial \bar\rho_+} 
+\bar\rho_+^2 \frac{\partial^2 f_0}{\partial \bar\rho_+^2}
= \frac{1+39\omega-4\omega^3}{36\omega}. \label{eqn:chi_exact}
\ee
For small couplings this can be expanded to yield
\be
\frac{\beta}{\bar\rho_+\chi_P} = 
1-\frac{\sqrt{3}}{4}\Gamma^{3/2} + \frac12\Gamma^3
-\frac{5}{2\sqrt{3}}\Gamma^{9/2} + \frac{21}{4}\Gamma^6 + 
{\cal O}(\Gamma^{13/2}).
\ee

\begin{figure}[ht]
\begin{center}
\leavevmode
\epsfxsize=0.63\textwidth
\epsfbox[15 20 580 450]{"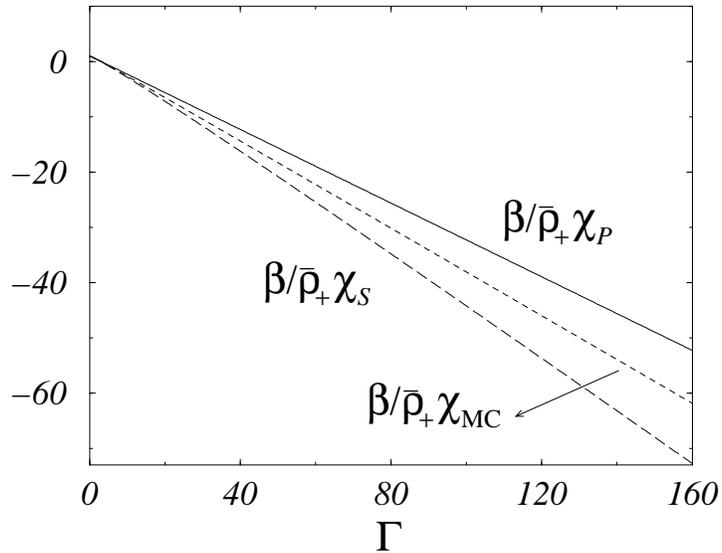"}
\end{center}
\caption{Comparison between the inverse compressibilities 
derived from the pressure, Eq.~(\protect\ref{eqn:chi_exact}), 
solid line $(\beta/\bar\rho_+\chi_P)$, and from the structure factor, 
Eq.~(\protect\ref{eqn:chi_S}), long-dashed line $(\beta/\bar\rho_+\chi_S)$.
The dashed line $(\beta/\bar\rho_+\chi_{\rm MC})$ 
represents the fit of the MC data over the 
interval $1\leq\Gamma\leq 160$ \protect\cite{DeWitt}.
\label{fig:fig1}}
\vspace{\baselineskip}
\end{figure}

To see if our expression for  $S(\kb)$ is consistent 
with the sum rules,  we expand~(\ref{eqn:skGDH})
around $\alpha=0$, using the asymptotic form of the spherical Bessel
function of the first kind,  
\be
\lim_{\alpha\to 0}\, j_\ell(\alpha x)= \frac{(\alpha x)^\ell}{(2\ell+1)!!} 
= \frac{(\alpha x)^\ell}{1.3.5\ldots (2\ell+1)}.
\ee
It is evident that, up to order ${\cal O}(k^0)$, 
only the isotropic $(\ell=0)$ terms of~(\ref{eqn:skGDH})
contribute to the structure factor. We find 
\bea
S^{-1}(\kb) &=& \left(\frac{\kd}{k}\right)^2 + 
\frac{\beta}{\bar\rho_+\chi_S}+{\cal O}(k^2), \\
\frac{\beta}{\bar\rho_+\chi_S} &=& 
1-\frac1{12}\int\limits_1^\omega {\rm d}\omega_\lambda\,
\frac{(2\omega_\lambda+1)(2\omega_\lambda^2-\omega_\lambda+2)}
{\omega_\lambda^2+\omega_\lambda+1} \nonumber\\
&=&   
1-\frac16 \left[ 1-2\omega+\omega^2-\frac{\pi\sqrt{3}}{2}
+ \frac34\ln\left(\frac{\omega^2+\omega+1}{3}\right) 
+\frac{3\sqrt{3}}{2}\tan^{-1}\left(\frac{2\omega+1}{\sqrt{3}}\right)
\right] . \label{eqn:chi_S}
\eea
In the low-density limit the inverse compressibility derived from 
the structure factor can be expanded to yield
\be
\frac{\beta}{\bar\rho_+\chi_S} =
1-\frac{\sqrt{3}}{4}\Gamma^{3/2} + \frac12\Gamma^3
-\frac{5}{2\sqrt{3}}\Gamma^{9/2} + \frac{81}{16}\Gamma^6 + 
{\cal O}(\Gamma^{13/2}).
\ee
We see that the structure factor satisfies \textit{exactly}
the charge-neutrality and the second-moment conditions, while the 
compressibility sum rule is satisfied only to order $\Gamma^{9/2}$. This
results from the fact that, in order to simplify the calculations,
we have neglected the dependence of the cavity size and shape on the imposed 
density variation $(\Delta)$. Clearly, if this was taken into account,
the theory would be completely internally self consistent. 
Nevertheless, even at this level
of approximation, the lack of self consistency is quite small over 
the full range of relevant coupling constants, as can be measured by the 
inverse compressibilities derived from the thermodynamic $(\chi_P)$ 
and the structure factor $(\chi_S)$ routes,
Eqs.~(\ref{eqn:chi_exact})~and~(\ref{eqn:chi_S}), 
respectively; see Fig.~(\ref{fig:fig1}).
The fit of the MC data \cite{DeWitt} leads to an inverse
compressibility $(\chi_{\rm MC})$ which is between the two previous ones.   

\begin{figure}[ht]
\begin{center}
\leavevmode
\epsfxsize=0.63\textwidth
\epsfbox[15 20 580 450]{"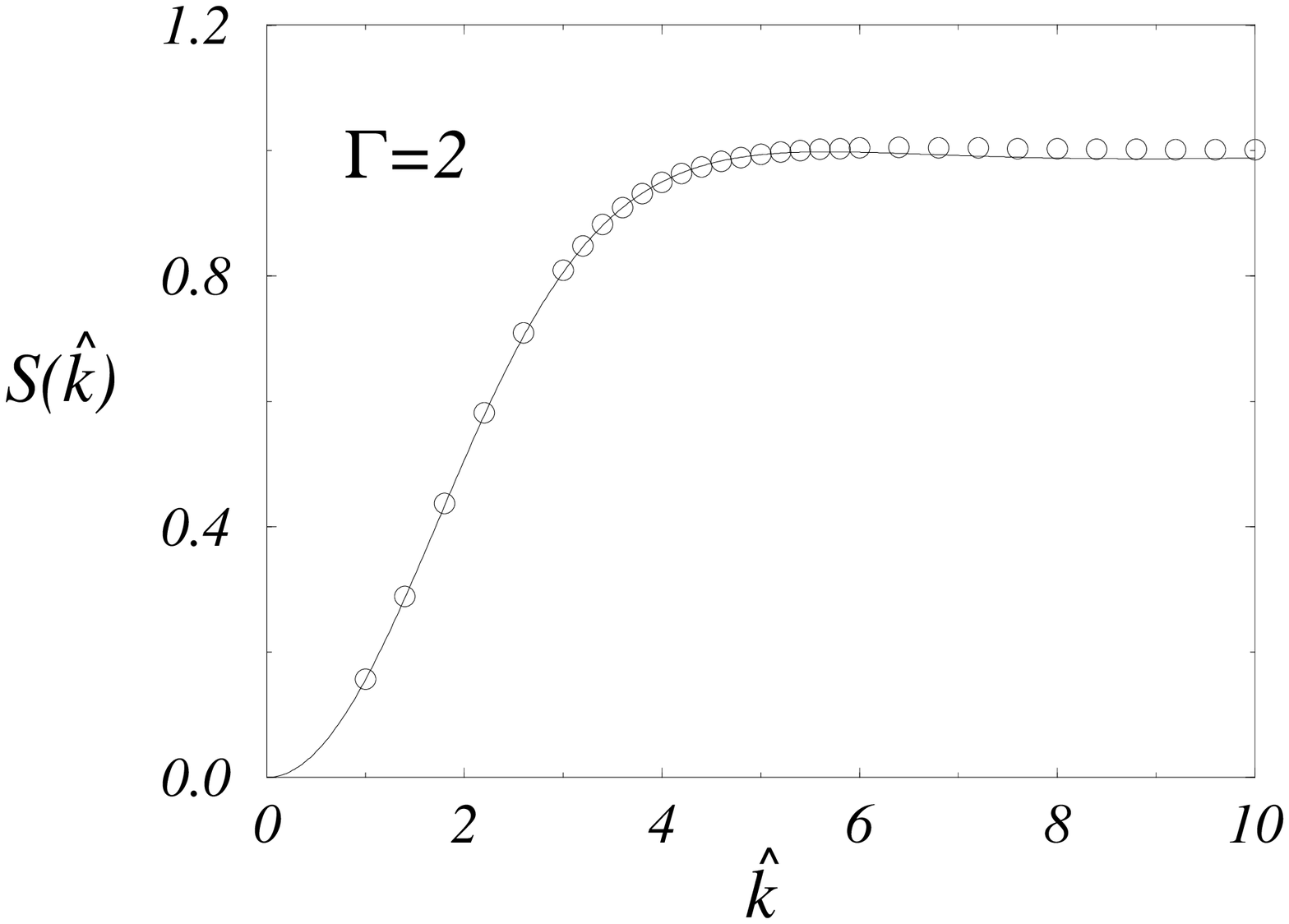"}
\end{center}
\caption{Structure factor $S(\hat{k})$
for $\Gamma=2$. The solid line is our 
expression~(\protect\ref{eqn:skGDH}) 
calculated up to $\ell=6$, while the circles
represent the MC data \protect\cite{Galam}.
\label{figure:fig2}}
\begin{center}
\leavevmode
\epsfxsize=0.63\textwidth
\epsfbox[15 20 580 450]{"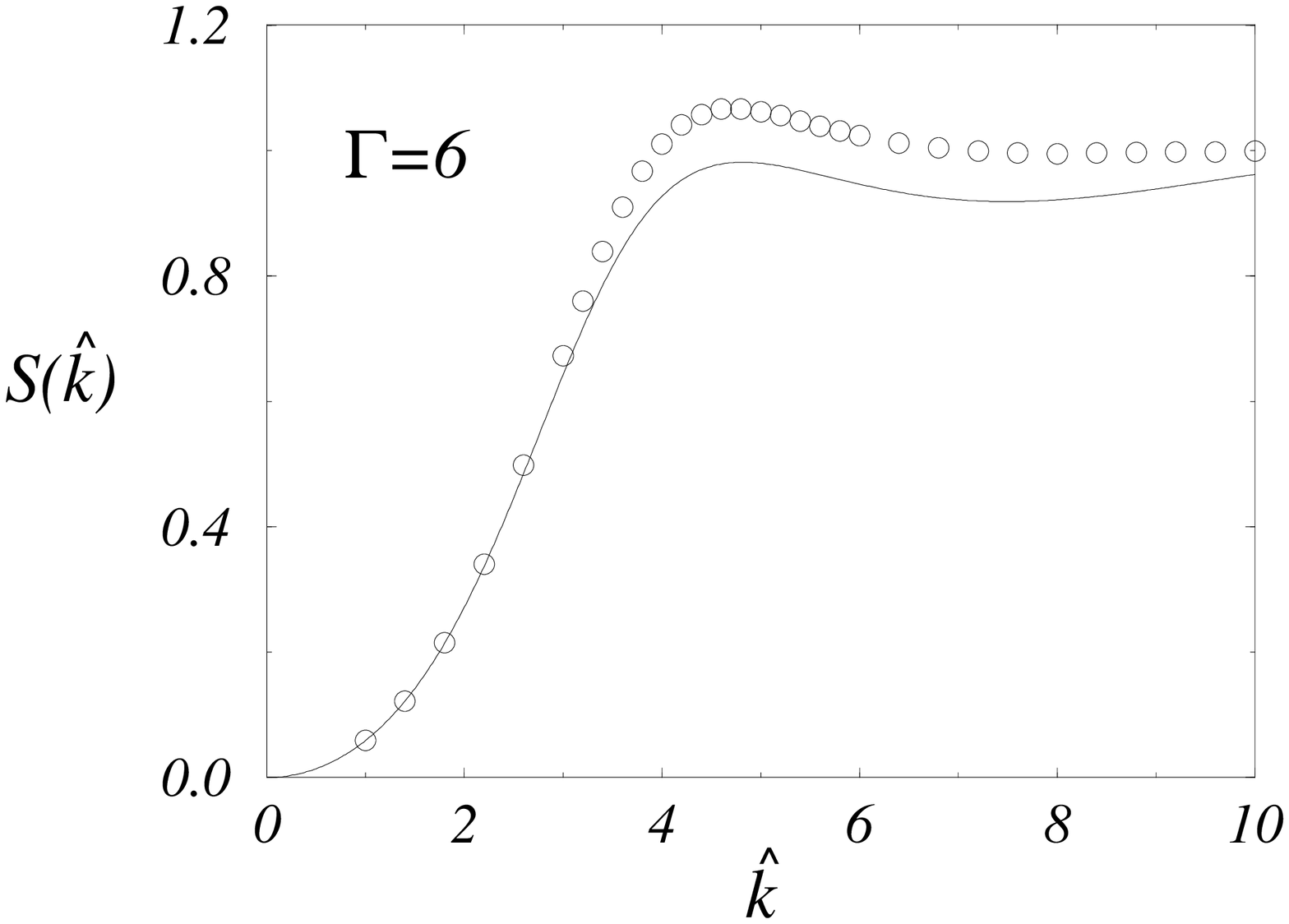"}
\end{center}
\caption{Structure factor $S(\hat{k})$
for $\Gamma=6$. The solid line is our 
expression~(\protect\ref{eqn:skGDH}) 
calculated up to $\ell=6$, while the circles
represent the MC data \protect\cite{Galam}.
\label{figure:fig3}}
\vspace{\baselineskip}
\end{figure}

\begin{figure}[ht]
\begin{center}
\leavevmode
\epsfxsize=0.63\textwidth
\epsfbox[15 20 580 450]{"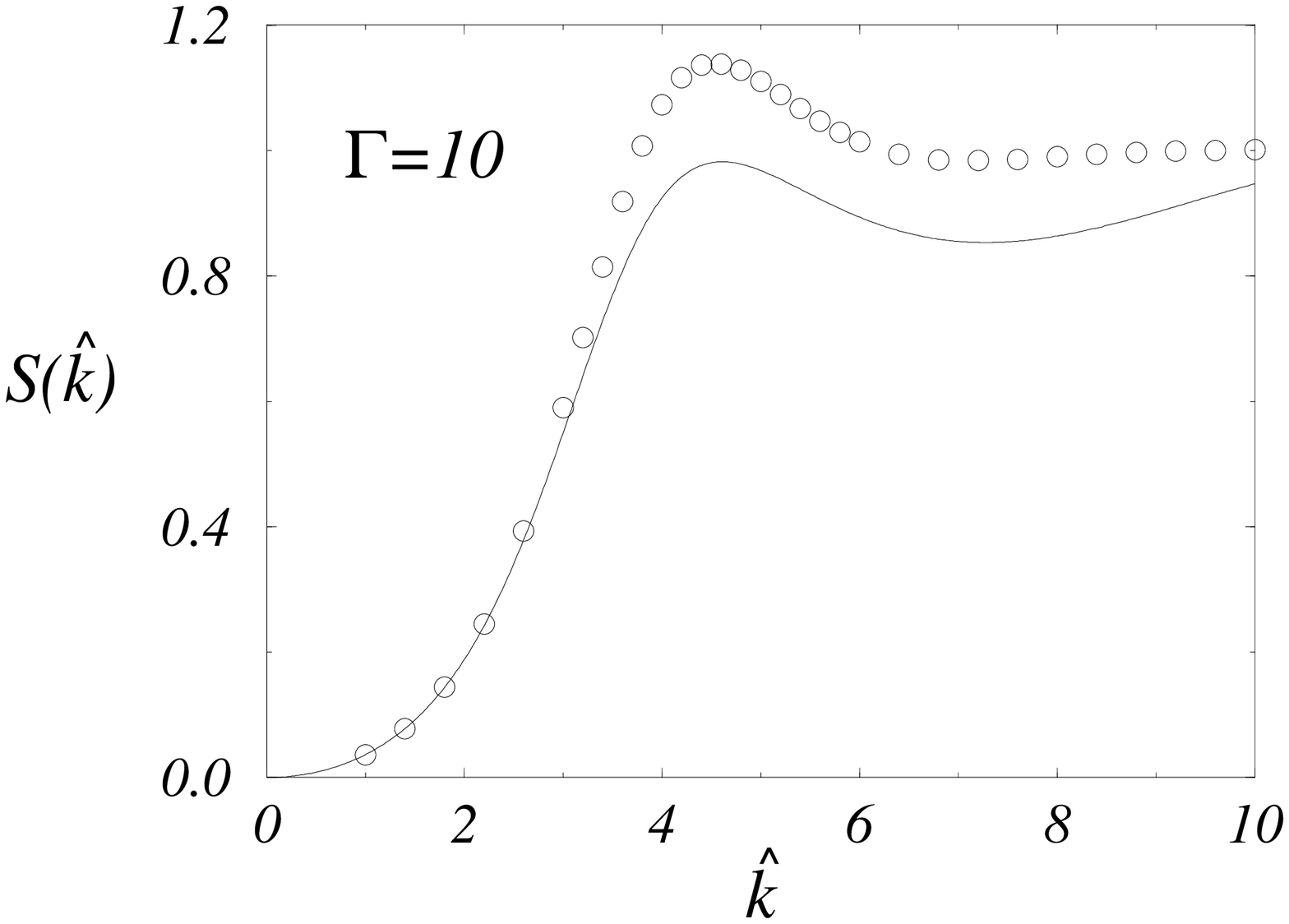"}
\end{center}
\caption{Structure factor $S(\hat{k})$
for $\Gamma=10$. The solid line is our 
expression~(\protect\ref{eqn:skGDH}) 
calculated up to $\ell=6$, while the circles
represent the MC data \protect\cite{Galam}.
\label{figure:fig4}}
\begin{center}
\leavevmode
\epsfxsize=0.63\textwidth
\epsfbox[15 20 580 450]{"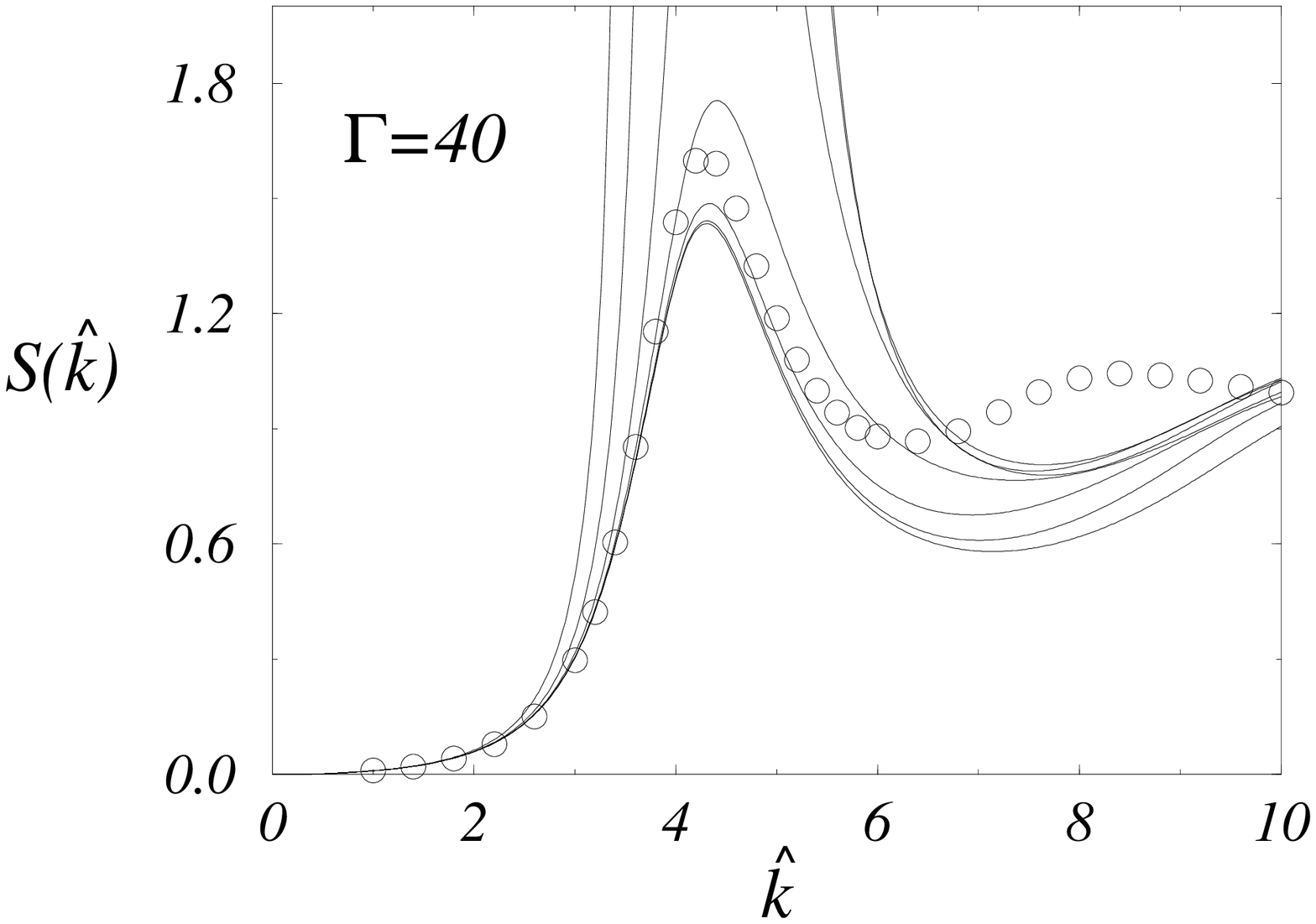"}
\end{center}
\caption{Structure factor $S(\hat{k})$
for $\Gamma=40$. The solid lines are our 
expression~(\protect\ref{eqn:skGDH}) with the $\ell=0,1,2,\ldots$ 
up to the $\ell=6$ terms 
included in the sum (from top to bottom). The circles represent 
the MC data \protect\cite{Galam}. We note that for 
higher values of $\Gamma$, more and more terms will need to
be included in order to achieve convergence in the vicinity of 
the first peak. 
\label{figure:fig5}}
\vspace{\baselineskip}
\end{figure}

Defining $\hat{k} \equiv kd=\alpha \sqrt{3\Gamma}$,  
we have explicitly carried out the summation for the first six terms
of the infinite series in Eq.~(\ref{eqn:skGDH}). The results for the 
structure factor for various values of $\Gamma$, obtained
\textit{without any fitting parameters,} are plotted in 
Figs.~(\ref{figure:fig2}) to~(\ref{figure:fig5}). The agreement with the
MC simulations \cite{Galam}
is quite encouraging. We should note, however, that for
higher couplings, in the vicinity of the first peak, the series is 
slowly convergent. This is the reason why we did not attempt to carry 
the calculations for $\Gamma>40$.

\section{Conclusions}

We have presented the generalized Debye-H\"uckel theory of the
one-component plasma. The linearity of the theory allows for explicit 
calculations of all the thermodynamic functions, as well as the structure
factor, which is expressed as an infinite series.
The linearity also insures the internal consistency of the 
theory. The agreement with the Monte Carlo simulations, obtained 
\textit{without any fitting parameters,} is quite good,
suggesting that the existence of an effective cavity surrounding each ion 
is theoretically justified.

\ack
 
This work has been supported by the Brazilian agencies CNPq (Conselho 
Nacional de Desenvolvimento Cient\'{\i}fico e Tecnol\'ogico), 
CAPES (Coordena\c{c}\~ao de Aperfei\c{c}oamento de Pessoal de
N\'{\i}vel Superior)
and
FAPERGS (Funda\c{c}\~ao de Amparo \`a Pesquisa do Estado do 
Rio Grande do
Sul).

\appendix

\section{Some useful integrals}

In this appendix we present some integrals which appear along 
the text. We introduce the Kronecker delta,
\be
\delta_{{\bms{k}}{\bf 0}} = 
{\displaystyle\frac{1}{V}}(2\pi)^3 \delta^3(\kb),
\ee
where $V=(2\pi)^3 \delta^3({\bf 0})$
is the volume of the system and 
$\delta^3(\kb)$ is the three-dimensional Dirac delta function,
\be
\delta^3(\kb) \equiv \left(\frac{1}{2\pi}\right)^3 
\int {\rm d}^3\rb'\,\exp\left({\rm i}\kb\cdot\rb'\right).
\label{eqn:diracdelta}
\ee

From the definition~(\ref{eqn:diracdelta})
of the three-dimensional Dirac delta function, it follows directly
\bea
\int {\rm d}^3\rb'\, \cos\kb\cdot\rb'&=&
\frac12 \int {\rm d}^3\rb'\, \left[
\exp\left({\rm i}\kb\cdot\rb'\right)
+\exp\left(-{\rm i}\kb\cdot\rb'\right)\right]\nonumber\\
&=& \frac12 \left(2\pi\right)^3 \left[\delta^3(\kb)+ \delta^3(-\kb)\right]
=(2\pi)^3 \delta^3(\kb)
=V \delta_{{\bms{k}}{\bf 0}} ,\label{eqn:coskr} \\  
\int {\rm d}^3\rb'\, \cos^2\kb\cdot\rb'&=&
\frac12 \int {\rm d}^3\rb'\, \left(1+\cos 2\kb\cdot\rb'\right) =
\frac12\left[ V+ (2\pi)^3 \delta^3(2\kb) \right] \nonumber\\
&=&
\frac{V}2 \left(1+\delta_{{\bms{k}}{\bf 0}}\right).
\label{eqn:coskr2}
\eea
A simple generalization of~(\ref{eqn:coskr2}) leads to
\be
\int {\rm d}^3\rb'\, \cos^2\left(\kb\cdot\rb'+
\ell\frac{\pi}{2}\right)=
\frac{V}2 \left[1+ (-1)^\ell \delta_{{\bms{k}}{\bf 0}}\right],
\ee
where $\ell$ is an integer.

Expressing $1/r$ as the inverse Fourier transform 
\be
\frac1r = \frac{1}{2\pi^2}\int {\rm d}^3\kb\, \frac1{k^2}
\exp\left(-{\rm i}\kb\cdot\rb\right),
\ee
we have 
\bea
\int {\rm d}^3\rb\, {\rm d}^3\rb'\, 
\frac{\cos \kb\cdot\rb\:  \cos\kb\cdot\rb'}{|\rb-\rb'|}
&=& \frac{1}{4\pi^2}\int {\rm d}^3\rb\, {\rm d}^3\rb'\, {\rm d}^3\qb\,
\frac{1}{q^2}\exp\left[-{\rm i}\qb\cdot\left(\rb-\rb'\right)\right] 
\nonumber\\
&&\hspace{0.1\textwidth}\mbox{}\!\times
\left[\cos\kb\cdot\left(\rb+\rb'\right) 
+ \cos\kb\cdot\left(\rb-\rb'\right) \right]
\nonumber\\
&=& \frac{1}{8\pi^2}\int {\rm d}^3\rb\, {\rm d}^3\rb'\, {\rm d}^3\qb\,
\frac{1}{q^2}
\sum_{\alpha_1=\pm}
\exp\left[{\rm i}\left(\alpha_1\kb-\qb\right)\cdot\rb\right]
\nonumber\\
&&\hspace{0.1\textwidth}\mbox{}\!\times
\sum_{\alpha_2=\pm}
\exp\left[{\rm i}\left(\alpha_2\kb+\qb\right)\cdot\rb'\right]\nonumber\\
&=& \frac12(2\pi)^4 \int {\rm d}^3\qb\, \frac1{q^2}
\sum_{\alpha_1,\alpha_2=\pm}
\delta^3\!\left(\alpha_1\kb-\qb\right)
\delta^3\!\left(\alpha_2\kb+\qb\right)
\nonumber\\
&=& \frac{(2\pi)^4}{k^2}\left[\delta^3({\bf 0})+
\frac12\delta^3(2\kb)+\frac12\delta^3(-2\kb) \right]\nonumber\\
&=& \frac{2\pi V}{k^2} \left( 1+\delta_{{\bms{k}}{\bf 0}} \right).
\label{eqn:coskr_overr2}
\eea

\section{Variation of the free-energy density}

In this appendix we obtain the variation of the 
reduced free-energy density, 
$\delta f=\beta\delta F/V$,  
up to quadratic order in the perturbation parameter $\Delta$.

The Helmholtz free energy $F$ is written as a functional of the
mobile-ion density $\rho_{+}(\rb)$. 
The variation $\delta F$ is obtained using the functional Taylor series,
\bea
\beta\delta F\left[\rho_{+}(\rb)\right] &=&
\beta F\left[\rho_{+}(\rb)\right]-\beta F\left[\bar\rho_{+}\right]
\nonumber\\
&=& \int {\rm d}^3\rb'\,\left.\frac{\beta\delta F}{\delta\rho_+(\rb')}
\right|_{\rho_{+}(\bms{r})=\bar\rho_{+}}
\delta\rho_+(\rb') \nonumber\\
&&\mbox{}\!
+ \frac12 \int {\rm d}^3\rb'\,{\rm d}^3\rb''
\left.\frac{\beta\delta^2 F}{\delta\rho_+(\rb')\,\delta\rho_+(\rb'')}
\right|_{\rho_{+}(\bms{r})=\bar\rho_{+}}
\delta\rho_+(\rb')\,\delta\rho_+(\rb''). 
\eea

The linear term can be written as 
\be 
\beta\delta F^{(1)}\left[\rho_{+}(\rb)\right] =
 \int {\rm d}^3\rb'\left.\frac{\beta\delta F}{\delta\rho_+(\rb')}
\right|_{\rho_{+}(\bms{r})=\bar\rho_{+}}\delta\rho_+(\rb')
=
\int {\rm d}^3\rb'\,\beta \mu(\rb')\,\delta\rho_+(\rb'), 
\ee
where $\mu(\rb)$ is the chemical potential at the position $\rb$,
\be
\mu(\rb)\equiv\left.\frac{\delta F}{\delta\rho_+(\rb)}
\right|_{\rho_{+}(\bms{r})=\bar\rho_{+}} .
\label{eqn:chemicalpot}
\ee
However, at the thermodynamical equilibrium, the chemical potential
of the system is constant and is independent of position, 
\be
\mu(\rb) = \bar\mu, \qquad\qquad \forall\rb.
\ee
Using the imposed variation~(\ref{eqn:rho_variation}) of the 
mobile-ion density,
\be
\delta \rho_+(\rb) = \bar\rho_+ \Delta\cos\kb\cdot\rb =
\frac12 \bar\rho_+ \Delta
\left[\exp\left({\rm i}\kb\cdot\rb\right)
+\exp\left(-{\rm i}\kb\cdot\rb\right) \right] , 
\label{eqn:deltarho}
\ee
the linear term can be expressed as
\be
\beta\delta F^{(1)}\left[\rho_{+}(\rb)\right] =
\beta\bar\mu\bar\rho_+\Delta
\int {\rm d}^3\rb'\,\cos \kb\cdot\rb'
= \beta\bar\mu\bar\rho_+V\Delta\delta_{{\bms{k}}{\bf 0}} .
\ee

The quadratic term,
\bea
\beta\delta F^{(2)}\left[\rho_{+}(\rb)\right]&=&
\frac12 \int {\rm d}^3\rb'\,{\rm d}^3\rb''
\left.\frac{\beta\delta^2 F}{\delta\rho_+(\rb')\,\delta\rho_+(\rb'')}
\right|_{\rho_{+}(\bms{r})=\bar\rho_{+}}
\delta\rho_+(\rb')\,\delta\rho_+(\rb'') \nonumber\\
&=& \frac12 \int {\rm d}^3\rb'\,{\rm d}^3\rb''
\left[ \frac{\delta\left(\rb'-\rb''\right)}{\bar\rho_+}
- C_{++}\left(\rb'-\rb''\right) \right] 
\delta\rho_+(\rb')\,\delta\rho_+(\rb''),  
\eea
can be split into two parts, the ideal-gas contribution,
\be
\beta\delta F_{\rm ideal}^{(2)}
\left[\rho_{+}(\rb)\right] =
\frac1{2\bar\rho_+}
\int {\rm d}^3\rb'\,{\rm d}^3\rb''\,
\delta(\rb'-\rb'')\,
\delta\rho_+(\rb')\,\delta\rho_+(\rb''),
\ee
and the electrostatic contribution,
\be
\beta\delta F_{\rm elect}^{(2)}
\left[\rho_{+}(\rb)\right] 
=
-\frac12 \int {\rm d}^3\rb'\,{\rm d}^3\rb''\,
C_{++}\left(\rb'-\rb''\right) 
\delta\rho_+(\rb')\,\delta\rho_+(\rb''),
\label{eqn:f_var_app}
\ee
where $C_{++}\left(\rb'-\rb''\right)$ is the direct 
correlation function.  

Using~(\ref{eqn:coskr2}) and~(\ref{eqn:deltarho}),
the ideal-gas contribution can be straightforwardly obtained,
\bea
\beta\delta F_{\rm ideal}^{(2)}\left[\rho_{+}(\rb)\right] &=&
\frac1{2\bar\rho_+}
\int {\rm d}^3\rb' \left[\delta\rho_+(\rb')\right]^2
=\frac12\bar\rho_+\Delta^2 \int {\rm d}^3\rb' \,\cos^2 \kb\cdot\rb'
\nonumber \\
&=& \frac{V}{4} \bar\rho_+\Delta^2
\left(1+\delta_{{\bms{k}}{\bf 0}}\right).
\eea

To evaluate the electrostatic contribution we
use~(\ref{eqn:deltarho}), and express 
$C_{++}\left(\rb'-\rb''\right)$
as the inverse Fourier transform 
\be
{C}_{++}(\rb) =
\left(\frac{1}{2\pi}\right)^3
\int {\rm d}^3\kb\; \hat{C}_{++}(\kb) \exp 
\left(-{\rm i} \kb\cdot\rb\right),
\ee
leading to  
\bea
\beta\delta F_{\rm elect}^{(2)}\left[\rho_{+}(\rb)\right] &=&
-\frac18\bar\rho_+^2\Delta^2 \left(\frac{1}{2\pi}\right)^3
\int {\rm d}^3\rb'\,{\rm d}^3\rb''\,{\rm d}^3\qb
\; \hat{C}_{++}(\qb)
\nonumber\\
&&\qquad\mbox{}\times 
\sum_{\alpha_1=\pm}
\exp\left[{\rm i}\left(\alpha_1\kb-\qb\right)\cdot\rb'\right]
\sum_{\alpha_2=\pm}
\exp\left[{\rm i}\left(\alpha_2\kb+\qb\right)\cdot\rb''\right] \nonumber\\
&=&- \frac18\bar\rho_+^2\Delta^2 \left(2\pi\right)^3  
\int {\rm d}^3\qb\; \hat{C}_{++}(\qb)
\sum_{\alpha_1,\alpha_2=\pm}
\delta^3\!\left(\alpha_1\kb-\qb\right)
\delta^3\!\left(\alpha_2\kb+\qb\right) \nonumber\\
&=&-\frac18\bar\rho_+^2\Delta^2 \left(2\pi\right)^3
\left\{
\hat{C}_{++}(\kb)\left[ 
\delta^3\!\left({\bf 0}\right)+\delta^3\!\left(2\kb\right)\right]
+ \hat{C}_{++}(-\kb)\left[ 
\delta^3\!\left({\bf 0}\right)+\delta^3\!\left(-2\kb\right)\right]
\right\}\nonumber\\
&=& -\frac{V}4\hat{C}_{++}(\kb)\bar\rho_+^2\Delta^2
\left(1+\delta_{{\bms{k}}{\bf 0}}\right) ,
\eea
where we have used the symmetry of the direct
correlation function, 
$\hat{C}_{++}(\kb)=\hat{C}_{++}(-\kb)$.

Combining all the pieces, the variation of the reduced free-energy density 
can be written as 
\bea
\delta f\left[\rho_{+}(\rb)\right] &=&
\frac{\beta}V\delta F^{(1)}\left[\rho_{+}(\rb)\right]+ 
\frac{\beta}V\delta F_{\rm ideal}^{(2)}\left[\rho_{+}(\rb)\right] +
\frac{\beta}V\delta F_{\rm elect}^{(2)}\left[\rho_{+}(\rb)\right] 
\nonumber\\
&=&
\beta\bar\mu\bar\rho_+\Delta\delta_{{\bms{k}}{\bf 0}} +
\frac14\left[1-\bar\rho_+\hat{C}_{++}(\kb)\right]
\bar\rho_+\Delta^2
\left(1+\delta_{{\bms{k}}{\bf 0}}\right)\nonumber\\
&=&
\beta\bar\mu\bar\rho_+\Delta\delta_{{\bms{k}}{\bf 0}} +
\frac14{S}^{-1}(\kb)\bar\rho_+\Delta^2 
\left(1+\delta_{{\bms{k}}{\bf 0}}\right) ,
\eea
where $S(\kb)$ is the structure factor. 
Note that the linear contribution to the variation has a 
Kronecker delta $(\delta_{{\bms{k}}{\bf 0}})$ factor,
which expresses the translational invariance~(\ref{eqn:chemicalpot}) of 
the equilibrium chemical potential $\bar\mu$ of the system.

\section{Green's function associated with the induced potential}

In this appendix we obtain the Green's function associated
with the differential equation satisfied by the
induced potential $\tilde\phi(\rb,\rb')$.

The Green's function $G\left(\Rb,\Rb'\right)$
associated with~(\ref{eqn:induced_potential1}), 
where $\Rb=\rb-\rb'$, satisfies the homogeneous equation \cite{Lee}
\bea
\left[ \nabla_{\!R}^2-\kd^2 \Theta\left(|\Rb|-h\right) \right]
G \left(\Rb,\Rb'\right) &=& -4\pi\delta^3(\Rb-\Rb') \nonumber\\
&=& 
-\frac{4\pi}{R^2}\delta(R-R')\,\delta(\cos\theta-\cos\theta')\,
\delta(\varphi-\varphi') 
, \label{eqn:green_eq}
\eea
with $\Theta(\xi)$ the Heaviside step function.

The general solution of~(\ref{eqn:green_eq}) can be written as \cite{Lee}
\be
G \left(\Rb,\Rb'\right) =  \kd 
\sum_{\ell=0}^\infty G_\ell\left(\kd R,\kd R'\right)
P_\ell \left(\frac{\Rb\cdot\Rb'}{RR'}\right), \label{eqn:green_exp}
\ee
where $P_\ell(\xi)$ denotes a Legendre polynomial. 

Replacing~(\ref{eqn:green_exp}) into~(\ref{eqn:green_eq}),
multiplying both sides by 
$\displaystyle P_{\ell'} \left(\frac{\Rb\cdot\Rb'}{RR'}\right)$ and 
integrating over the angular coordinates $\theta$  and $\varphi$,
we obtain the equation satisfied by the radial functions 
$G_\ell\left(s,s'\right)$,
\be
\left[\frac{{\rm d}^2}{{\rm d}s^2}+\frac{2}{s}\frac{{\rm d}}{{\rm d}s} 
- \Theta\left(s-x\right)
-\frac{\ell(\ell+1)}{s^2} \right] G_\ell\left(s,s'\right) =
-\frac{1}{s^2}\left(2\ell+1\right)\delta(s-s'), \label{eqn:radialeqn}
\ee
where we have introduced the adimensional 
variables $s=\kd R$, $s'=\kd R'$ and $x=\kd h$.
To obtain~(\ref{eqn:radialeqn}) we have used the
property of the Dirac delta function, 
\be
\delta(\kd R-\kd R')=\frac1\kd \delta(R-R'),
\ee
the addition theorem for the Legendre polynomials,
\bea
P_\ell \left(\frac{\Rb\cdot\Rb'}{RR'}\right) &=& 
P_\ell \left[\cos\theta\cos\theta' 
+\sin\theta\sin\theta'\cos(\varphi-\varphi')
\right]\\
&=&
P_\ell(\cos\theta)
P_\ell(\cos\theta')
+2\sum_{m=1}^\ell \frac{(\ell-m)!}{(\ell+m)!}\,
P_\ell^m\! \left(\cos\theta\right)
P_\ell^m\! \left(\cos\theta'\right)\cos 
m(\varphi-\varphi'), \nonumber
\eea
and their orthogonality, 
\be
\int\limits_{-1}^{1} {\rm d}(\cos\theta)
\, P_\ell(\cos\theta)\,P_{\ell'}(\cos\theta) = \frac2{2\ell+1}
\delta_{\ell\ell'}.
\ee

The solutions of~(\ref{eqn:radialeqn}) that are finite for $s\to 0$ and 
vanish as $s\to\infty$ can be written as 
\be
G_\ell\left(s,s'\right) =
\left\{
\begin{array}{ll}
A_{11} s^\ell, & \qquad\mbox{ for }0<s<s'<x,\\
A_{12} s^\ell+ A_{13}s^{-(\ell+1)}, & \qquad\mbox{ for }0<s'<s<x,\\
A_{21} s^\ell, & \qquad\mbox{ for }0<s<x<s',\\
A_{22} k_\ell(s), & \qquad\mbox{ for }0<s'<x<s,\\
A_{31} k_\ell(s), & \qquad\mbox{ for }0<x<s'<s,\\
A_{32} i_\ell(s) + A_{33} k_\ell(s), & \qquad\mbox{ for }0<x<s<s',
\end{array}
\right.
\label{eqn:greensolution}
\ee
where the coefficients $\{A_{mn}\}$ are functions of $x$ and $s'$
to be determined by the boundary conditions;
$i_\ell(s)$ and $k_\ell(s)$
are the modified spherical Bessel functions of the first and the 
third kinds \cite{Bateman}, respectively,
\bea
i_\ell(s)&=& \sqrt{\frac{\pi}{2 s}}\, I_{\ell+1/2} (s), \\
k_\ell(s)&=& \sqrt{\frac{2}{\pi s}}\, K_{\ell+1/2} (s). 
\eea

Using the symmetry property of the Green's function \cite{Jackson},
we can rewrite Eqs.~(\ref{eqn:greensolution}) as
\bea
G_\ell^{(1)}\left(s,s'\right) &=& A_{1}s^\ell s'^\ell + 
{B}_{1}\frac{s_<^\ell}{s_>^{\ell+1}},\qquad \mbox{ for } 0<s,s'< x,\\
G_\ell^{(2)}\left(s,s'\right) &=& A_{2}s_<^\ell k_\ell(s_>),
\qquad \mbox{ for } 0<s_< < x < s_>,\\
G_\ell^{(3)}\left(s,s'\right) &=& A_{3} k_\ell(s) k_\ell(s')
+ B_3 i_\ell(s_<) k_\ell(s_>),\qquad \mbox{ for } 0<x<s,s',
\eea
where $s_<=\min(s,s')$, $s_>=\max(s,s')$,
and the coefficients $\{A_n,B_n\}$ depend now only on the size of the 
exclusion hole $x$.

The coefficients $\{B_n\}, n=1,3$, are obtained by imposing the 
discontinuity of the derivative of $G_\ell(s,s')$ associated 
with the Dirac delta function,
\bea
\frac{\rm d}{{\rm d}s}
\left[ sG_\ell^{(n)}\left(s,s'\right)\right]_{s=s'+\epsilon} &-&
\frac{\rm d}{{\rm d}s}
\left[ sG_\ell^{(n)}\left(s,s'\right)\right]_{s=s'-\epsilon}
= -\frac{2\ell+1}{s'},
\eea
where $\epsilon$ is a positive infinitesimal.
Using the Wronskian of the modified spherical Bessel 
functions \cite{Bateman},
\be
W[k_\ell(s),i_\ell(s)]=
k_\ell(s) i'_\ell(s)-i_\ell(s) k'_\ell(s)= \frac1{s^2},
\ee
this leads to  
\bea
B_1&=& 1, \\
B_3&=& 2\ell+1.
\eea

The coefficients $\{A_n\}, n=1,2,3$, are obtained by imposing the 
continuity of $G_\ell(s,s')$ and of its derivative 
across the spherical surface at $s=s'=x$,
\bea
\left.G_\ell^{(1)}\left(s,s'\right)\right|_{s=s'=x} &=&
\left.G_\ell^{(2)}\left(s,s'\right) \right|_{s=s'=x}= 
\left.G_\ell^{(3)}\left(s,s'\right)\right|_{s=s'=x}, \\
\left.\frac{\rm d}{{\rm d}s}G_\ell^{(1)}\left(s,s'\right)
\right|_{s=s'+\epsilon=x}  &=&
\left.\frac{\rm d}{{\rm d}s}G_\ell^{(2)}\left(s,s'\right)
 \right|_{s=s'+\epsilon=x} ,
\eea
and using the following relations of the modified spherical
Bessel functions \cite{Bateman} 
to express $i_\ell(x)$, $k_\ell(x)$ and $k_\ell'(x)$
in terms of $i_{\ell\pm 1}(x)$ and $k_{\ell\pm 1}(x)$,
\bea
\frac1{x^2}&=& i_{\ell+1}(x) k_\ell(x) +k_{\ell+1}(x) i_\ell(x)  ,\\
\left(2\ell+1\right) k_\ell(x) &=& x k_{\ell+1}(x)-x k_{\ell-1}(x), \\
-\left(2\ell+1\right) k_\ell'(x) &=& \ell k_{\ell-1}(x)+ 
\left(\ell+1\right) k_{\ell+1}(x),
\eea
which yield
\bea
A_1&=& -\frac{k_{\ell-1}(x)}{ x^{2\ell+1} k_{\ell+1}(x)} ,\\
A_2&=&\frac{2\ell+1}{x^{\ell+2} k_{\ell+1}(x) },\\
A_3&=&\left(2\ell+1\right) \frac{ i_{\ell+1}(x)}{ k_{\ell+1}(x)}.
\eea

Therefore, the Green's function $G \left(\Rb,\Rb'\right)$
is given by the expansion~(\ref{eqn:green_exp}), with the
radial functions $G_\ell\left(s,s'\right)$ defined by \cite{Lee} 
\bea
G_\ell^{(1)}\left(s,s'\right) &=& 
\frac{s_<^\ell}{s_>^{\ell+1}}-
\frac{s^\ell s'^\ell k_{\ell-1}(x)}{x^{2\ell+1} k_{\ell+1}(x)}
,\quad \mbox{ for } 0<s,s' < x,\\
G_\ell^{(2)}\left(s,s'\right) &=& \left(2\ell+1\right)
\frac{s_<^\ell k_\ell(s_>)}{x^{\ell+2} k_{\ell+1}(x)},
\quad \mbox{ for } 0<s_< < x < s_>,\\
G_\ell^{(3)}\left(s,s'\right) &=& \left(2\ell+1\right)
\left[ \frac{ i_{\ell+1}(x)}{ k_{\ell+1}(x)}
 k_\ell(s) k_\ell(s')+ i_\ell(s_<) k_\ell(s_>)
\right] ,\quad \mbox{ for } 0<x<s,s'.
\eea

\section{The perturbative solution of the induced potential}

In this appendix we obtain the induced potential 
$\tilde\phi(\rb,\rb')$ recursively,
up to order $\Delta^2$, at the center of the exclusion hole, 
$|\rb-\rb'|=0$.
 
Let us obtain the induced potential \textit{outside} the
exclusion hole, $|\rb-\rb'|\ge h$, which we will denote by
$\tilde\phi_>(\rb,\rb')$. Clearly, this potential is produced
by the charge distribution inside and outside the cavity. Let us
first calculate the contribution to the potential arising from the 
charge \textit{inside} the hole, $\tilde\phi^{(<)}_>(\rb,\rb')$.
Since our final goal is to calculate the potential at the center of
the cavity to order $\Delta^2$, it is sufficient to calculate the induced 
potential \textit{outside} the hole to order $\Delta$, 
see Eq.~(\ref{eqn:nonhomogeneity}).
Using the Green's function $G\left(\Rb,\Rb'\right)$ derived in 
appendix C, where $\Rb=\rb-\rb'$, we find to first order in $\Delta$,
\bea
\tilde\phi_>^{(<)}(\rb,\rb')&=&
\tilde\phi_>^{(<)}(\Rb+\rb',\rb')=
\frac{1}{D}\int\limits_{|\bms{R}'|\le h} {\rm d}^3 \Rb'\,
\varrho\left(\Rb'\right) G\left(\Rb,\Rb'\right) \nonumber\\
&=& \frac{q\kd}{D} \int\limits_{|\bms{R}'|\le h} {\rm d}^3 \Rb'
\left\{\delta^3(\Rb') - 
\bar\rho_{+}\left[1+\Delta\cos\kb\cdot(\Rb'+\rb')\right] \right\} \nonumber\\
&&\mbox{}\!\times\sum_{\ell=0}^\infty
P_\ell \left(\frac{\Rb\cdot\Rb'}{RR'}\right) \nonumber\\
&=& \frac{1}{\beta q}\left(\lambda_{\rm B}\kd-\frac13x^3\right)
\frac{k_0(\kd R)}{x^2k_1(x)} 
-\frac{q\bar\rho_+\kd}{D}\Delta\,
{\rm Re}\left[\exp\left({\rm i}\kb\cdot\rb' \right)
\sum_{\ell=0}^\infty \left(2\ell+1\right) 
\right.\nonumber\\
&&\left.\mbox{}\!\times \frac{k_\ell(\kd R)}{x^{\ell+2} k_{\ell+1}(x)}
\int\limits_{|\bms{R}'|\le h} {\rm d}^3 \Rb'
\left(\kd R'\right)^\ell \exp\left({\rm i}\kb\cdot\Rb' \right)
P_\ell \left(\frac{\Rb\cdot\Rb'}{RR'}\right) \right] ,
\label{eqn:phi1_>}
\eea
recalling that $x=\kd h$.
The first term of~(\ref{eqn:phi1_>}) can be simplified 
using the identities
\bea
k_1(x)&=&\left(1+x\right) \frac{{\rm e}^{-x}}{x^2}, 
\label{eqn:identity1}\\
\lambda_{\rm B}\kd &=& \frac13 \left[\left(1+x\right)^3-1\right]
= x \left(1+x\right) +\frac13 x^3. \label{eqn:identity2}
\eea
The relation~(\ref{eqn:identity2}) is the defining equation for the
cavity size, $x$, Eq.~(\ref{eqn:ocp_hole}). 
It is important to remember that it does not take 
into account the imposed variation in the ionic density, and as result 
will be responsible for the violation of the compressibility sum rule.

To simplify the second term of~(\ref{eqn:phi1_>}), we note first that,
without loss of generality, we can choose the $z$ axis along the 
$\kb$ direction,
\bea
\cos\theta &=& \frac{\kb\cdot\Rb}{kR}, 
\qquad\qquad
\cos\theta'= \frac{\kb\cdot\Rb'}{kR'},\\
\tan \varphi &=&\frac{\Rb\cdot \hat{\bm{y}}}{\Rb\cdot \hat{\bm{x}}},
\qquad\qquad
\tan \varphi' =\frac{\Rb'\cdot \hat{\bm{y}}}{\Rb'\cdot \hat{\bm{x}}} ,
\eea
so that the addition theorem for the Legendre polynomials
can be written as
\bea
P_\ell \left(\frac{\Rb\cdot\Rb'}{RR'}\right) &=& 
P_\ell \left[\cos\theta\cos\theta' 
+\sin\theta\sin\theta'\cos(\varphi-\varphi')
\right]\\
&=&
P_\ell (\cos\theta)
P_\ell (\cos\theta')
+2\sum_{m=1}^\ell \frac{(\ell-m)!}{(\ell+m)!}\,
P_\ell^m (\cos\theta)
P_\ell^m (\cos\theta')\cos 
m(\varphi-\varphi'). \nonumber
\eea
Performing the integrations over the azimuthal angle $\varphi'$,
only the $m=0$ terms survive, 
\bea
\tilde\phi_>^{(<)}(\rb,\rb')
&=& \frac{1}{\beta q}
x{\rm e}^x k_0(\kd R)  
-\frac{\kd^3}{\beta q}\Delta\,
{\rm Re}\left[\exp\left({\rm i}\kb\cdot\rb' \right)
\sum_{\ell=0}^\infty \left(2\ell+1\right)
\frac{k_\ell(\kd R)}{x^{\ell+2} k_{\ell+1}(x)}
\right.\nonumber\\
&&\left.\mbox{}\!\times  P_\ell (\cos\theta) 
\int\limits_0^h {\rm d}R'\,R'^2 
\left(\kd R'\right)^\ell 
\int\limits_{-1}^{1} {\rm d}(\cos\theta')
 \exp\left({\rm i}k R' \cos\theta' \right) P_\ell(\cos\theta')
\right].
\eea
To proceed, we use the plane-wave expansion,
\be
\exp\left({\rm i}k R' \cos\theta' \right) = 
\sum_{\ell=0}^\infty \left(2\ell+1\right) {\rm i}^\ell  j_\ell(kR')
P_\ell (\cos\theta'), 
\ee
where  $j_\ell(\xi)$ is the spherical Bessel function
of the first kind, 
\be
j_\ell(\xi) = \sqrt{\frac{\pi}{2 \xi}}\, J_{\ell+1/2} (\xi).
\ee
The integrations over the polar angle $\theta'$ and over the radial 
coordinate $R'$ can be performed using the 
orthogonality of the Legendre polynomials,  
\be
\int\limits_{-1}^{1} {\rm d}(\cos\theta')
\, P_\ell(\cos\theta')\,P_{\ell'}(\cos\theta') = \frac2{2\ell+1}
\delta_{\ell\ell'},
\ee
and the recursion relation for the spherical Bessel function of the 
first kind, 
\be
\frac{\rm d}{{\rm d}\xi} \left[\xi^{\ell+2}j_{\ell+1}(\xi)\right] 
=  \xi^{\ell+2}j_{\ell}(\xi) ,
\ee
which yields
\bea
\tilde\phi_>^{(<)}(\rb,\rb')
&=& \tilde\phi_>^{(<)}(\Rb+\rb',\rb')
=
\frac{1}{\beta q}
x{\rm e}^x k_0(\kd R) -\frac{1}{\alpha\beta q}\Delta 
\sum_{\ell=0}^\infty \left(2\ell+1\right)
\cos\left(\kb\cdot\rb'+\ell\frac{\pi}{2}\right) \nonumber\\
&&\mbox{}\!\times 
\frac{j_{\ell+1}(\alpha x)}{k_{\ell+1}(x)} k_\ell(\kd R)\, 
P_\ell (\cos\theta),
\label{eqn:phioutside1}
\eea
where $\alpha=k/\kd$.

Substituting $\tilde\phi_>^{(<)}(\rb,\rb')$ into the expression 
for the charge density \textit{outside} the exclusion hole, 
Eq.~(\ref{eqn:nonhomogeneity}), we can now calculate the contribution 
to the potential \textit{outside} the exclusion hole 
arising from the \textit{external} charge, 
$\tilde\phi_>^{(>)}(\rb,\rb')$. To order $\Delta$ we find  
\bea
\tilde\phi_>^{(>)}(\rb,\rb')&=&
\tilde\phi_>^{(>)}(\Rb+\rb',\rb')
= \frac{1}{D}\int\limits_{|\bms{R}'|\ge h} {\rm d}^3 \Rb'\,
\varrho\left(\Rb'\right) G\left(\Rb,\Rb'\right) \nonumber\\ 
&=&
-\frac{\kd^3}{4\pi}\Delta \int\limits_{|\bms{R}'|\ge h} {\rm d}^3 \Rb'\,
\tilde\phi_>^{(<)}(\Rb'+\rb',\rb')
\cos\kb\cdot\left(\Rb'+\rb'\right)
\nonumber\\
&&\mbox{}\!\times
\sum_{\ell=0}^\infty G_\ell^{(3)} (\kd R,\kd R')\,
P_\ell \left(\frac{\Rb\cdot\Rb'}{RR'}\right) \nonumber\\
&=&
- \frac{1}{\beta q}x{\rm e}^x \Delta
\sum_{\ell=0}^\infty \left(2\ell+1\right)
\cos\left(\kb\cdot\rb'+\ell\frac{\pi}{2}\right) \Xi_\ell(\kd R,\alpha)\,
P_\ell (\cos\theta),
\eea
where the function $\Xi_\ell(s,\alpha)$ is defined by 
\bea
\Xi_\ell(s,\alpha)&=&
\int\limits_x^\infty {\rm d}s'\,s'^2\,  
k_0(s')\,\frac{G_\ell^{(3)}(s,s')}{2\ell+1}\, j_\ell(\alpha s') \nonumber\\
&=&
\frac{i_{\ell+1}(x)}{k_{\ell+1}(x)} k_\ell(s)
\int\limits_x^\infty {\rm d}\xi\, \xi^2 k_0(\xi)k_\ell(\xi)
j_\ell(\alpha\xi)+ k_\ell(s) \int\limits_x^{s} {\rm d}\xi\, 
\xi^2 k_0(\xi)i_\ell(\xi)j_\ell(\alpha\xi) \nonumber\\
&&\mbox{}\!+ i_\ell(s) \int\limits_{s}^\infty {\rm d}\xi\,
 \xi^2 k_0(\xi)k_\ell(\xi)j_\ell(\alpha\xi).
\eea

We are now able to find the induced potential \textit{inside} the
exclusion hole, $|\rb-\rb'|\le h$, up to order $\Delta^2$,
\bea
\tilde\phi_<(\rb,\rb')&=&
\tilde\phi_<(\Rb+\rb',\rb')=
\frac{1}{D}\int {\rm d}^3 \Rb'\,
\varrho\left(\Rb'\right) G\left(\Rb,\Rb'\right)\nonumber\\
&=& \frac{q\kd}{D} \int\limits_{|\bms{R}'|\le h} {\rm d}^3 \Rb'
\left\{\delta^3(\Rb') - 
\bar\rho_{+}\left[1+\Delta\cos\kb\cdot(\Rb'+\rb')\right] \right\} \nonumber\\
&&\mbox{}\!\times\sum_{\ell=0}^\infty G_\ell^{(1)} (\kd R,\kd R')\,
P_\ell \left(\frac{\Rb\cdot\Rb'}{RR'}\right) 
-\frac{\kd^3}{4\pi}\Delta 
\int\limits_{|\bms{R}'|\ge h} {\rm d}^3 \Rb'\,
\tilde\phi_> (\Rb'+\rb',\rb')\nonumber\\
&&\mbox{}\!\times\cos \kb\cdot\left(\Rb'+\rb'\right) 
\sum_{\ell=0}^\infty G_\ell^{(2)} (\kd R,\kd R')\,
P_\ell \left(\frac{\Rb\cdot\Rb'}{RR'}\right),
\label{eqn:inducedinside}
\eea
where the induced potential \textit{outside} the
exclusion hole, up to order $\Delta$, is given by
\be
\tilde\phi_> (\rb,\rb') = 
\tilde\phi_>^{(<)} (\rb,\rb') + \tilde\phi_>^{(>)} (\rb,\rb').
\ee
Since we need just the mean induced electrostatic potential
$\psi(\rb')$ felt by the positive ion fixed at
$\rb'$, that is, at the center of the exclusion hole,
$\Rb={\bf 0}$, and recalling that $G_\ell(\kd R=0,\kd R')=0, 
\forall \ell>0$,  only the isotropic $(\ell=0)$ terms
of~(\ref{eqn:inducedinside}) contribute, 
\bea
\psi(\rb')&=& \lim_{{\bms{R}}\to{\bf 0}}
\left[\tilde\phi_<(\Rb+\rb',\rb')
-\frac{q}{DR}\right] \nonumber\\
&=& 
 \lim_{R\to 0}\left[
\frac{q\kd}{D} \int\limits_{|\bms{R}'|\le h} {\rm d}^3 \Rb'\,
G_0^{(1)}(\kd R,\kd R')\,\delta^3(\Rb')-\frac{q}{DR} \right]
\nonumber\\
&&\mbox{}\!-
\frac{q\bar\rho_+\kd}{D} \int\limits_{|\bms{R}'|\le h} {\rm d}^3 \Rb'
\,G_0^{(1)}(0,\kd R')\, \left[1+\Delta\cos\kb\cdot(\Rb'+\rb')\right]
\nonumber\\
&& \mbox{}\!
 -\frac{\kd^3}{4\pi}\Delta 
\int\limits_{|\bms{R}'|\ge h} {\rm d}^3 \Rb'\,
G_0^{(2)}(0,\kd R')\,
\tilde\phi_> (\Rb'+\rb',\rb') \cos \kb\cdot\left(\Rb'+\rb'\right)
\nonumber\\
&=& 
 \lim_{R\to 0}\left[
\frac{q\kd}{D}G_0^{(1)}(\kd R,0)-\frac{q}{DR} \right]
-\frac{\kd^3}{\beta q}\int\limits_0^h {\rm d}R'\,R'^2G_0^{(1)}
(0,\kd R')\nonumber\\
&&\mbox{}\!
-\frac{\kd^3}{\beta q}\Delta\cos\kb\cdot\rb'
\int\limits_0^h {\rm d}R'\,R'^2G_0^{(1)}
(0,\kd R') j_0(k R')
\nonumber\\
&&\mbox{}\!
 -\frac{\kd^3}{4\pi}\frac{1}{x^2 k_1(x)}\Delta
\int\limits_{|\bms{R}'|\ge h} {\rm d}^3 \Rb'\,
k_0(\kd R')\,\tilde\phi_> (\Rb'+\rb',\rb')
\cos \kb\cdot\left(\Rb'+\rb'\right).
\label{eqn:induced_at_center}
\eea
Using the explicit form of $G_0^{(1)}(s,s')$ and 
$\tilde\phi_> (\rb,\rb')$,
and performing the angular integrations, we  
obtain
\bea
\beta q\psi(\rb')&=&
-\lambda_{\rm B}\kd \frac{k_{-1}(x)}{xk_1(x)}
-\int\limits_0^x {\rm d}s\,s^2\left[\frac1s-
\frac{k_{-1}(x)}{xk_1(x)}\right] \nonumber\\
&&\mbox{}\!
-\Delta\cos\kb\cdot\rb' \left\{
\int\limits_0^x {\rm d}s\,s^2\left[\frac1s-
\frac{k_{-1}(x)}{xk_1(x)}\right] j_0(\alpha s)
 + \frac{x {\rm e}^{2x}}{1+x} 
\int\limits_x^\infty {\rm d}s\,s^2 k_0^2(s)j_0(\alpha s) \right\}
\nonumber\\
&&\mbox{}\!
+ \frac{{\rm e}^x}{\alpha\left(1+x\right)}\Delta^2 
\sum_{\ell=0}^\infty \left(2\ell+1\right)
\cos^2\left(\kb\cdot\rb'+\ell\frac{\pi}{2}\right)  
\frac{j_{\ell+1}(\alpha x)}{k_{\ell+1}(x)} 
\int\limits_x^\infty {\rm d}s\,s^2 k_0(s)k_\ell(s)j_\ell(\alpha s)
\nonumber\\
&&\mbox{}\!
+ \frac{x{\rm e}^{2x}}{1+x}\Delta^2 
\sum_{\ell=0}^\infty \left(2\ell+1\right)
\cos^2\left(\kb\cdot\rb'+\ell\frac{\pi}{2}\right)
\int\limits_x^\infty {\rm d}s\,s^2 k_0(s)\Xi_\ell(s,\alpha)j_\ell(\alpha s).
\label{eqn:beta_psi1}
\eea
The first terms of~(\ref{eqn:beta_psi1}) can be simplified 
using~(\ref{eqn:identity1}) and~(\ref{eqn:identity2}), 
supplemented by the identities
\bea
k_0(x)&=& \frac{{\rm e}^{-x}}{x}, \\
\frac{k_{-1}(x)}{k_1(x)} &=& \frac{k_1(x)-k_0(x)/x}{k_1(x)}= 
\frac{x}{1+x}, \\
j_0(\xi)&=& \frac{\sin \xi}{\xi}. 
\eea
Furthermore, 
expressing $i_\ell(\xi)$ in terms of $k_\ell(\xi)$ using the 
relation \cite{Bateman}
\be
i_\ell(\xi)= -\frac12\left[ k_\ell(-\xi)+(-1)^\ell k_\ell(\xi)\right], 
\ee
and defining the $\ell$-th grade polynomial $g_\ell(\xi)$ associated 
with the modified spherical
Bessel function of the third kind 
$k_\ell(\xi)$ by the identity \cite{Bateman}
\be
g_\ell(\xi)\equiv {\rm e}^{\xi}{\xi^{\ell+1}}k_\ell(\xi)
= \sum_{m=0}^\ell\frac{\Gamma(\ell+m+1)}{ 2^m m!\,
\Gamma(\ell-m+1)}\,\xi^{\ell-m} = 
\sum_{m=0}^\ell
\frac{(2m)!}{2^m m!}\, {{\ell+m}\choose{2m}}\,\xi^{\ell-m} ,
\ee
where $\Gamma(m)=(m-1)!$ is the Euler gamma function,  
it is possible to express the last integral of~(\ref{eqn:beta_psi1}), 
which is two-dimensional,
in terms of one-dimensional quadratures,  
\bea
\int\limits_x^\infty {\rm d}s\,s^2 k_0(s)\Xi_\ell(s,\alpha)j_\ell(\alpha s)
&=&{\rm e}^{-2x}
(-1)^{\ell+1}\left\{\frac12\frac{g_{\ell+1}(-x)}{g_{\ell+1}(x)}
\left[{\cal I}_\ell^+(x,\alpha)\right]^2\right.\nonumber\\
&&\hspace{0.2\textwidth}
\left.\mbox{}\!\vphantom{\frac{g_\ell}{g_\ell}}+
{\cal I}_\ell^+(x,\alpha){\cal I}_\ell^-(x,\alpha)
-{\cal I}_\ell^0(x,\alpha)\right\},
\eea
where $\{{\cal I}_\ell^\nu\}, \nu=\pm,0,$ are the 
one-dimensional integrals
\bea
{\cal I}_\ell^-(s,\alpha) &=& 
\int\limits_0^{s} {\rm d}\xi\, \xi^{-\ell} g_\ell(-\xi)
j_\ell(\alpha \xi), \\
{\cal I}_\ell^0(x,\alpha) &=& 
\int\limits_x^\infty {\rm d}s\, s^{-\ell} g_\ell(s)
j_\ell(\alpha s){\cal I}_\ell^-(s,\alpha)\exp\left[2(x-s)\right], \\
{\cal I}_\ell^+(x,\alpha) &=& 
\int\limits_x^\infty {\rm d}s\, s^{-\ell} g_\ell(s)
j_\ell(\alpha s)\exp\left[2(x-s)\right].
\eea
We stress that the functions $\{{\cal I}_\ell^0(x,\alpha)\}$ 
represent one-dimensional quadratures,
since the integrals $\{{\cal I}_\ell^-(s,\alpha)\}$ can be
expressed in explicit form,
for all values of $\ell$, 
in terms of trigonometric functions and of the sine integral,
\be
{\rm Si\,}(t) = \int\limits_0^t {\rm d}\xi\, \frac{\sin \xi}{\xi}.
\ee
To illustrate, we give 
the three first integrals:
\bea
{\cal I}_0^-(s,\alpha) &=& \frac{{\rm Si\,}(\alpha s)}{\alpha}, \\
{\cal I}_1^-(s,\alpha) &=&-\frac{1}{\alpha}+\frac{\cos\alpha s}
{2\alpha s}+
\left(\frac1{\alpha^2 s}
-\frac1{2\alpha^2 s^2}\right)
\sin\alpha s+\frac12{{\rm Si\,}(\alpha s)}, \\
{\cal I}_2^-(s,\alpha) &=&
-1+\left(\frac{3}{2 \alpha^2 s}-\frac{3}{\alpha^2 s^2}+
\frac{9}{4\alpha^2 s^3}+\frac{3}{8s} \right) \cos\alpha s\nonumber\\
&&\mbox{}\!+
\left(-\frac{3}{2\alpha^3 s^2}+\frac{3}{\alpha^3 s^3}
-\frac{9}{4 \alpha^3 s^4}+\frac{3}{8 \alpha s^2}\right) \sin\alpha s+
\left(\frac1{2\alpha}+\frac{3\alpha}8 \right) {\rm Si\,}(\alpha s).
\eea 

Therefore, the final form for the 
mean induced electrostatic potential at the center of 
the exclusion hole (in unities of $\beta q$)
reads
\bea
\beta q\psi(\rb')&=& -\frac12x (x+2) 
-\Delta\cos\kb\cdot\rb' \left[\frac{1}{\alpha^2} 
-\frac{\sin \alpha x}{\alpha^3(1+x)}-
\frac{\cos \alpha x}{\alpha^2(1+x)} + \frac{x}{1+x}\,
{\cal I}_0^+(x,\alpha) \right]\nonumber\\
&&\mbox{}\!+ \frac{1}{\alpha(1+x)}
\Delta^2\sum_{\ell=0}^{\infty} \left(2\ell+1\right)
\cos^2\left(\kb\cdot\rb'+\ell\frac{\pi}{2}\right) 
\frac{x^{\ell+2} j_{\ell+1}(\alpha x)}{g_{\ell+1}(x)}\,
{\cal I}_\ell^+(x,\alpha)\nonumber\\
&&\mbox{}\!+  \frac{x}{1+x} 
\Delta^2\sum_{\ell=0}^{\infty} \left(-1\right)^{\ell+1} \left(2\ell+1\right)
\cos^2\left(\kb\cdot\rb'+\ell\frac{\pi}{2}\right)\nonumber\\
&&\mbox{}\!\times
\left\{\frac12 \frac{g_{\ell+1}(-x)}{g_{\ell+1}(x)}
\left[{\cal I}_\ell^+(x,\alpha)\right]^2+
{\cal I}_\ell^+(x,\alpha){\cal I}_\ell^-(x,\alpha)
-{\cal I}_\ell^0(x,\alpha)\right\}.
\eea

\end{document}